# On the Origin and Implications of Li$_2$O$_2$ Toroid Formation in Nonaqueous Li-O$_2$ Batteries

Nagaphani B Aetukuri[1*], Bryan D McCloskey[1,2,3], Jeannette M. García[1], Leslie E Krupp[1], Venkatasubramanian Viswanathan[4*] and Alan C. Luntz[1,5]

*1 IBM Almaden Research Center, San Jose, CA, 95120*
*2 Department of Chemical and Biomolecular Engineering, University of California, Berkeley, CA, 94720*
*3 Environmental Energy Technologies Division, Lawrence Berkeley National Laboratory, Berkeley, CA, 94720*
*4 Department of Mechanical Engineering, Carnegie Mellon University, Pittsburgh, PA, 15213*
*5 SUNCAT, SLAC National Accelerator Laboratory, Menlo Park, CA 94025*

*Address correspondence to: phani@alumni.stanford.edu (N.B.A); venkvis@cmu.edu (V.V)*

**The lithium-air (Li-O$_2$) battery has received enormous attention as a possible alternative to current state-of-the-art rechargeable Li-ion batteries given their high theoretical specific energy. However, the maximum discharge capacity in nonaqueous Li-O$_2$ batteries is limited to a small fraction of its theoretical value due to the insulating nature of lithium peroxide (Li$_2$O$_2$), the battery's primary discharge product. In this work, we show that the inclusion of trace amounts of electrolyte additives, such as H$_2$O, significantly improve the capacity of the Li-O$_2$ battery. These additives trigger a solution-based growth mechanism due to their solvating properties, thereby circumventing the Li$_2$O$_2$ conductivity limitation. Experimental observations and a growth model imply that this solution mechanism is responsible for Li$_2$O$_2$ toroid formation. We present a general formalism describing an additive's tendency to trigger the solution process, providing a rational design route for electrolytes that afford larger Li-air battery capacities.**

The recent surge in activity seeking batteries with energy densities surpassing that possible with Li-ion intercalation technology is fueled by the goal of developing mass-market electrification of road transportation. The nonaqueous Li-air battery has attracted the most attention to date because of its very high theoretical specific energy[1-2]. In this battery, the net electrochemical reaction is 2 Li + O$_2$ ⇆ Li$_2$O$_2$, with the forward reaction describing discharge of the battery and the reverse describing charge[3-8]. The high theoretical specific energy arises because Li metal is used as the anode and ambient air can act as a source for $O_2$. At present, there are still substantial technical obstacles to developing a practical Li-air battery[9-10]. Perhaps the most significant challenges arise from parasitic chemistry and electrochemistry during battery cycling[11-18] and the electrical passivation of the cathode that occurs during discharge [19-22]. The first limits rechargeability while the second limits capacity to less than theoretically possible, especially at higher current densities, implying a poor capacity-power tradeoff in the battery[21]. The electrical passivation is caused by the build up of $Li_2O_2$, a wide band gap insulator during discharge[3,19-23]. This inhibits charge transfer from the cathode to the $Li_2O_2$ – electrolyte interface where the discharge electrochemistry occurs.

In ethereal electrolytes, many authors[24-29] have reported that large $Li_2O_2$ toroids of variable sizes (100 nm – 1 µm) are produced during discharge at low currents. However, during the course of our studies with nearly anhydrous ethereal electrolytes, we have never observed toroid formation at any current rate, apparently only forming thin conformal coatings of $Li_2O_2$ on the cathode surface. Understanding the origin of the large toroid features and ultimately controlling the morphology of $Li_2O_2$ growth during discharge is extremely critical for attaining high discharge capacities since these toroids circumvent the $Li_2O_2$ charge transport limitations.

In this article, we combine experimental measurements with theoretical modeling to show that there are two possible paths for $Li_2O_2$ growth on the cathode. One involves a surface electrochemical mechanism previously described[8,30] that produces conformal $Li_2O_2$ coatings on the cathode surface whose thicknesses are limited by charge transport through $Li_2O_2$. We propose that the second path is a solution-mediated electrochemical process driven by $LiO_2$ partial solubility, where $O_2^-$ acts as a redox mediator and ultimately promotes the growth of $Li_2O_2$ toroids at low currents. Although solution-mediated processes have previously been suggested for the growth of $Li_2O_2$ [4-5,24,28] neither the electrochemical origins nor the conditions that favor the growth of large toroids have been outlined. In ethereal solvents, we demonstrate that the toroidal growth, and a concomitant discharge capacity increase, is only observed when trace quantities of $H_2O$ are added to the electrolyte. However, the presence of water also introduces unfavorable consequences for the overall battery electrochemistry. As a result, understanding how $H_2O$ enhances the solubility of $LiO_2$ allows us to develop insight as to which solvents/additives may induce solution-growth of $Li_2O_2$ (thereby enhancing the capacity of Li-$O_2$ batteries), but without the drawbacks that added $H_2O$ introduces.

Figure 1 presents scanning electron microscopy (SEM) images of Vulcan® XC72 carbon cathodes (XC72) extracted from Li-$O_2$ batteries after galvanostatic discharge to 1 mAh capacity at a current of 50 µA for 500-4000 ppm $H_2O$ added to the electrolyte (1 M Li-TFSI in dimethoxyethane, DME). The Li-$O_2$ batteries are identical to those used for differential electrochemical mass spectrometry (DEMS) experiments discussed previously[11-12] and procedures are described in the methods section of the supplementary information (SI). With no added $H_2O$, the XC72 cathode after discharge is indistinguishable from the pristine cathode prior to discharge. This suggests that the $Li_2O_2$ is only deposited as thin conformal films on the cathode surface. However, after addition of 500 ppm $H_2O$, small thin toroids (100-200 nm in size) become visible in the SEM images. As the $H_2O$ content increases, increasingly larger toroids are formed and layering within the toroids becomes more evident.

Figure 1 clearly shows that the existence of $Li_2O_2$ toroids in ether-based electrolytes, along with their shape, size and abundance, depends on the concentration of added $H_2O$ in the electrolyte. Previous reports[24-27,31] indicated toroids with a range of shapes and sizes in ethereal electrolytes and we suggest that this is likely due to varying levels of water contamination in the cells. Figure S1 in the SI shows a diminishment in toroid particle diameter with current at a fixed $H_2O$ content, with no toroids present at currents > 1 mA for 4000 ppm $H_2O$ concentration. At low $H_2O$ content, although toroids are still observed at very low currents, they disappear at currents much lower than 1 mA, in general agreement with prior observations[24,26,32]. We observe similar discharge morphology changes with increasing $H_2O$ content with other cathodes as well, e.g. TiC and AvCarb® P50 carbon paper (see Figs. S2 and S3 in SI). Toroid morphology is observed when comparable quantities of water are added to other electrolyte solvents, e.g. tetraethylene glycol dimethyl ether (TEGDME) and dimethylsulfoxide (DMSO) (see Figs. S4 and S5 in SI), thus, suggesting that the presence of $H_2O$ in the electrolyte is the dominant cause of the drastic $Li_2O_2$ morphology changes.

Additionally, discharge capacity increases as the $H_2O$ concentration increases (Fig. 2a), in agreement with previous studies[33]. At low discharge currents (50 μA), this is accompanied with an increase in toroid size. However, even at high currents (3 mA) where no toroid formation is apparent, an increase in discharge capacity is observed with water present (Fig. S10b). We argue that the improvement in capacity at both currents arises due to a solution-mediated mechanism for $Li_2O_2$ formation (discussed later) that overcomes charge transport limitations inherent in surface growth of $Li_2O_2$. While trace $H_2O$ has a positive impact on capacity, it is also critical to understand its effect on the battery chemistry and rechargeability.

Figure 3a shows X-ray diffractograms (XRD) near the $Li_2O_2$ (100) and (101) peaks from Avcarb P50 paper cathodes extracted from batteries otherwise similar to those studied in Fig 1. The only additional $H_2O$-induced XRD feature is a small peak at 30.65 degrees that has tentatively been identified as $Li_2NH$ (Fig. S6). These results confirm that the majority of the crystalline discharge product is $Li_2O_2$, regardless of electrolyte water content. Notably, no crystalline LiOH is observed in the XRD of the cathodes. The $Li_2O_2$ diffractograms clearly show a decreasing peak width as a function of increasing water content in the electrolyte solution, implying that the $Li_2O_2$ crystallite size increases, in agreement with the SEM images shown in Fig 1. Other authors have shown a decrease in XRD linewidth with current (presumably at fixed $H_2O$ content) as the toroid size increases[24,31,34]. For the nominally anhydrous battery, there was no apparent change in crystallite size with current (see Fig. 3b) Therefore, XRD provides no evidence for a transformation to an amorphous $Li_2O_2$ deposit at higher currents as suggested by others[24,35].

To quantify the effects of added $H_2O$ on the electrochemistry and possible, parasitic reactions, both quantitative DEMS and $Li_2O_2$ titration were employed [11-12]. If no parasitic electrochemistry (and chemistry) occurs during discharge, the expected yield of $Li_2O_2$ produced relative to the theoretical yield from the discharge capacity is unity, $Y_{Li2O2} = 1.0$. Likewise, two electrons are ideally utilized for each $O_2$ consumed, $(e^-/O_2)_{dis} = 2.00$, where $O_2$ consumption is monitored using a pressure decay measurement. Figure S7 and S8 in the SI show $(e^-/O_2)_{dis}$ and $Y_{Li2O2}$, respectively. The values without added $H_2O$ are consistent with previous measurements[12] for these low current conditions, i.e. $Y_{Li2O2} = 0.75$ for an XC72 cathode and $(e^-/O_2)_{dis} = 2.02$. However, both deviate further from their optimum values as $H_2O$ is added, demonstrating that the added $H_2O$ induces parasitic processes during discharge. We also note that a titration of both solvent and cathode yields a higher total peroxide content than that from the cathode itself, which we believe is indirect evidence for soluble $H_2O_2$ formation. It is possible that the additional parasitic chemistry is due to the formation of $H_2O_2$ and other soluble species, as will be discussed later.

Since $H_2O$ induces parasitic chemistries during discharge, it is not surprising that $H_2O$ impurities also affect the charging potential. Galvanostatic discharge-charge cycles as a function of $H_2O$ content are shown in Fig. S9 in the SI. These profiles demonstrate that although the initial charging potential $U_{chg}$ is nearly identical in all cases, the rate of increase in $U_{chg}$ with charging capacity $Q_{chg}$ is strongly dependent on the $H_2O$ content. We have previously suggested that the initial $U_{chg}$ is indicative of the low $Li_2O_2$ fundamental kinetic overpotentials[30], but that the increase in $U_{chg}$ with $Q_{chg}$ is related to the role of parasitic products in charging[36]. This interpretation for Fig. S9 in the SI is entirely consistent with the enhanced parasitic chemistry associated with added $H_2O$ (Figs. S7 and S8). Comparing the performance and (electro)chemistry of the anhydrous cells to those with trace amounts of $H_2O$ raises two questions: How does $H_2O$ increase capacity and induce toroid formation? And, is it possible to find another additive that can give the positive benefits of added $H_2O$ without its drawbacks?

Since the dimensions of the toroids are significantly larger than the charge transport-limited dimensions of 1-10 nm, we hypothesize that they must be formed by a solution-mediated mechanism that also contributes to the electrochemistry, as also suggested by others[24,29]. Thus, the net battery discharge is the sum of the two different contributions. Figure 4a shows discharge linear scan voltammograms (LSV), both with and without added $H_2O$ in DME. The anhydrous DME discharge LSV is similar to that observed previously[8] and was assigned to the surface process producing $Li_2O_2$ (the peak at ~2.5 V in Fig. 4a). The discharge LSV curve with 4000 ppm water (Fig. 4b) shows an additional peak at potentials lower than ~2.5 V. We suggest that this additional peak in the LSV is related to the solution-mediated growth of $Li_2O_2$. This additional peak represents an electrochemical process principally forming $Li_2O_2$, as indicated by a peroxide titration ($Y_{Li2O2}$ = 0.81 following the LSV with 4000 ppm added $H_2O$ in DME). Since current from both surface and solution processes is possible at typical galvanostatic conditions (where a potential plateau ranges from 2.4-2.7 V), both the surface and solution processes can contribute to galvanostatic discharges.

The schematic in Fig. 5 summarizes the two electrochemical paths for $Li_2O_2$ crystal growth. The surface electrochemical growth is given as before[30].

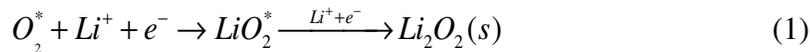

$$O_2^* + Li^+ + e^- \rightarrow LiO_2^* \xrightarrow{Li^+ + e^-} Li_2O_2(s) \qquad (1)$$

where * refers to a surface adsorbed species and (s) for solid. Both $Li^+ + e^-$ charge transfer or disproportionation ($2LiO_2^* \rightarrow Li_2O_2(s) + O_2$) can contribute to the second step of growth [8,37]. We also consider a second slower possible route for the growth of $Li_2O_2$ crystals induced by the generation of soluble reduced oxygen species in the presence of $H_2O$. The dominant $Li_2O_2$ equilibrium surface produced by reaction (1) is the O-rich (0001) surface, i.e. a $Li_2O_2$ surface with half a monolayer of $LiO_2$ adsorbed[23,30,38]. Its solubility is given as

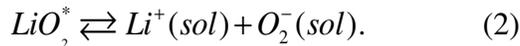

$$LiO_2^* \rightleftarrows Li^+(sol) + O_2^-(sol). \qquad (2)$$

The equilibrium is governed by the stability of the $Li^+$ and $O_2^-$ ions in solution (sol) relative to the $LiO_2^*$ adsorption energy on $Li_2O_2$. As described in detail in the SI, the stability of $Li^+$ and $O_2^-$ ions in solution is related to the Gutman Donor Number (DN) and acceptor number (AN), respectively and hence the LiO$_2$* solubility depends on these parameters. The addition of water triggers this dissolution process by solvating $O_2^-$ efficiently due to its very high AN of ~55. We believe that the solution soluble $O_2^-$ undergoes subsequent reaction on a growing Li$_2$O$_2$ toroid through the generic mechanism

$$2Li^+(sol) + 2O_2^-(sol) \rightleftharpoons Li_2O_2(s) + O_2(g) \qquad (3)$$

Many different detailed mechanisms could contribute to reaction (3). However, the key point is that Li$_2$O$_2$ solution growth uses $O_2^-(sol)$ as a redox shuttle. The most likely mechanism (shown in Fig. 5) is that: (a) LiO$_2$* solvates in an equilibrium fashion from the O-rich (0001) Li$_2$O$_2$ that conformally coats the cathode that was produced via the surface process (1) (since this is the dominant surface area of O-rich (0001) Li$_2$O$_2$ formed), (b) $O_2^-(sol)$ diffuses to a growing particle where it forms LiO$_2$* again, (c) two such LiO$_2$* disproportionate to form Li$_2$O$_2$ on a larger growing particle and (d) LiO$_2$* regenerates via reaction (1) at the empty site on the conformal layer. It is worth highlighting that the anhydrous DME has an AN ~ 10 and therefore does not induce enough solubility of LiO$_2$* to induce significant solution growth.

In a partially protic solvent (DME with added H$_2$O), $O_2^-(sol)$ is known to undergo disproportionation, ultimately forming H$_2$O$_2$. H$_2$O$_2$ formation is a relatively slow step in mixed aprotic-H$_2$O solvents[39]. However, this step, along with a reaction between H$_2$O and Li metal, slowly consumes the H$_2$O and eventually this reduction of water reduces the overall dissolution rate and ultimately terminates the solution growth mechanism.

Based on this mechanism, we developed an electrochemical model that accounts for the simultaneous surface and solution routes to formation of Li$_2$O$_2$, and where appropriate compare with the experiments above. This model also includes the disproportionation of $O_2^-(sol)$ in the presence of water. The model is described in detail in the SI. The evolution of the different chemical species, H$_2$O, $O_2^-(sol)$ and H$_2$O$_2$, during galvanostatic discharges for several different added H$_2$O contents is given in Fig.

S15. The significant increase in the concentration of $O_2^-(sol)$ in the presence of $H_2O$ promotes the solution mechanism in addition to the surface electrochemical growth of $Li_2O_2$ and therefore increases the discharge capacity (see Fig. 2b). However, the solution route ultimately shuts off due to a decrease in soluble $O_2^-(sol)$ as the $H_2O$ is consumed by conversion to $H_2O_2$ and by reaction with the Li anode. This consumption leads to a decreased solubility of $O_2^-(sol)$ which ultimately determines the maximum discharge capacity for a given $H_2O$ content. The maximum discharge capacity, therefore, is a function of the $H_2O$ content in the electrolyte. The results from the model are summarized in Fig 2b and are in qualitative agreement with Fig. 2a, i. e. a ~5 fold enhancement in discharge capacity for the 4000 ppm case relative to the anhydrous case. Using the same model, we also simulate the LSV for 4000 ppm added $H_2O$ in Fig. 4c. This clearly demonstrates the existence of two distinct peaks associated with the surface and solution electrochemical growth routes and is in good agreement with Fig. 4b. Note that the LSV experiments are shifted to lower potentials due to *iR* losses at these currents[20].

Clearly the solution mechanism allows particle sizes (and capacities) larger than the few nm dimensions defined by charge transport. How particles dynamically grow in such a process is a challenging problem. At low currents, large layered toroids are observed. At higher currents (and/or lower $H_2O$ content), much smaller particle sizes are formed. In the SI, we present a plausible kinetic model for the solution growth of these particles. It is based on assuming three coupled kinetic processes; a solution growth rate of $Li_2O_2$ on the particles, a passivation rate that helps terminate the solution growth (e. g. formation of $Li_2NH$ on the $Li_2O_2$ particle surface) and a rate for formation of defects/holes in the passivation layer. With reasonable kinetic rates and at low currents, layered toroids are naturally formed via this mechanism, with sizes dependent upon the $H_2O$ concentration and overall current. Two examples are given in Fig 5b and 5c, and an animation of the toroid growth is given in the SI. At higher currents, diffusion limitations of $O_2^-(sol)$ restrict the size of the growing particles, and at lower $H_2O$ concentrations, the shorter times (capacities) available for solution growth restrict the particle size (see Fig.

S1f). An extended discussion of this model and its relationship to the parameters of the solution mechanism are given in the SI.

While added $H_2O$ enhances discharge capacity in ether solvents by inducing a solution mechanism for growth of $Li_2O_2$, it also induces enhanced parasitic chemistry. A key question is whether other additives/solvents can also induce the solution growth mechanism, but perhaps without the additional parasitic chemistry. As discussed in detail in Section S3, the solubility of $LiO_2^*$ is determined by the Gutman DN and AN of the solvent, respectively. Based on experimental measurements on the redox potential shifts for $O_2/O_2^-$ and $Li/Li^+$, we develop an expression for the relative free energy of dissolution, Eq. (2), on any solvent (see section S3 for details). Fig. 6 shows a contour plot of the free energy of dissolution as a function of AN and DN with several known solvents labeled in the plot. DME and MeCN have limited propensity to solvate $LiO_2^*$ and hence are ineffective in promoting solution growth. Of the pure aprotic solvents, DMSO is relatively active because of its high DN. In fact, cathodic LSVs of cells employing DMSO-based electrolytes (see Fig. S12) also exhibit a weak second peak ascribed to the solution-mediated $Li_2O_2$ formation mechanism. Furthermore, very small toroids are observed on cathodes that are galavanostatically discharged in nearly anhydrous DMSO (~30 ppm water content, see Fig. S5a), at low currents, in agreement with Figure 6, but these grow substantially in size (and layering) with added $H_2O$. Another possible additive with strong solvating properties is $CH_3OH$. Addition of 4000 ppm $CH_3OH$ to DME does increase the maximum discharge capacity by ~3x relative to pure DME at 100 μA current (see Fig S10a). The LSV with methanol added to DME (see Fig. S11) also shows an additional peak (although the peak current is smaller compared to $H_2O$) that we attribute to the solution growth of $Li_2O_2$. Unfortunately, the pKa of protic $CH_3OH$ is slightly lower than $H_2O$, so a galvanostatic discharge also produces $H_2O_2$ and a $Y_{Li2O2}$ ~ 0.5 (0.4 mAh discharge capacity).

The enhanced discharge capacity and the growth of large $Li_2O_2$ toroids in the presence of added water is definitive evidence that the discharge capacity of Li-$O_2$ batteries need not be limited by the surface growth route and the electronically insulating nature of $Li_2O_2$. The solubility of $O_2^-$ in the battery electrolyte can activate a mechanism where $O_2^-$ acts as a redox mediator for the electrochemical growth of $Li_2O_2$ that is not limited by the charge transport of $Li_2O_2$. The few examples above show that this solution route is not limited to only added $H_2O$ and validates the analysis that lead to the predictions of Fig. 6. Solvents/additives that allow for increased solubility of $LiO_2^*$ will be the key for obtaining large discharge capacities, and we have developed a quantitative basis for the rational selection of solvents based on their acceptor and donor numbers. The desired additive must possess a high DN or/and AN while at the same time possess a high $pK_a$ to avoid $H_2O_2$ formation and related parasitic processes. The identification of such additives could pave the way towards enhancing the discharge capacity, while still optimizing the rechargeability of the $Li-O_2$ battery.

**AUTHOR CONTRIBUTIONS**

All authors contributed to the design of the research. N. B. A., J. M. G., and L. E. K. performed the experimental measurements, and N. B. A. performed the experimental data analysis. V. V. and A. C. L. designed the theoretical calculations, which V. V. then performed. N. B. A., B. D. M., V. V., and A. C. L. co-wrote the manuscript. All authors discussed the results and commented on the manuscript.

Competing financial interests: The authors declare no competing financial interests.

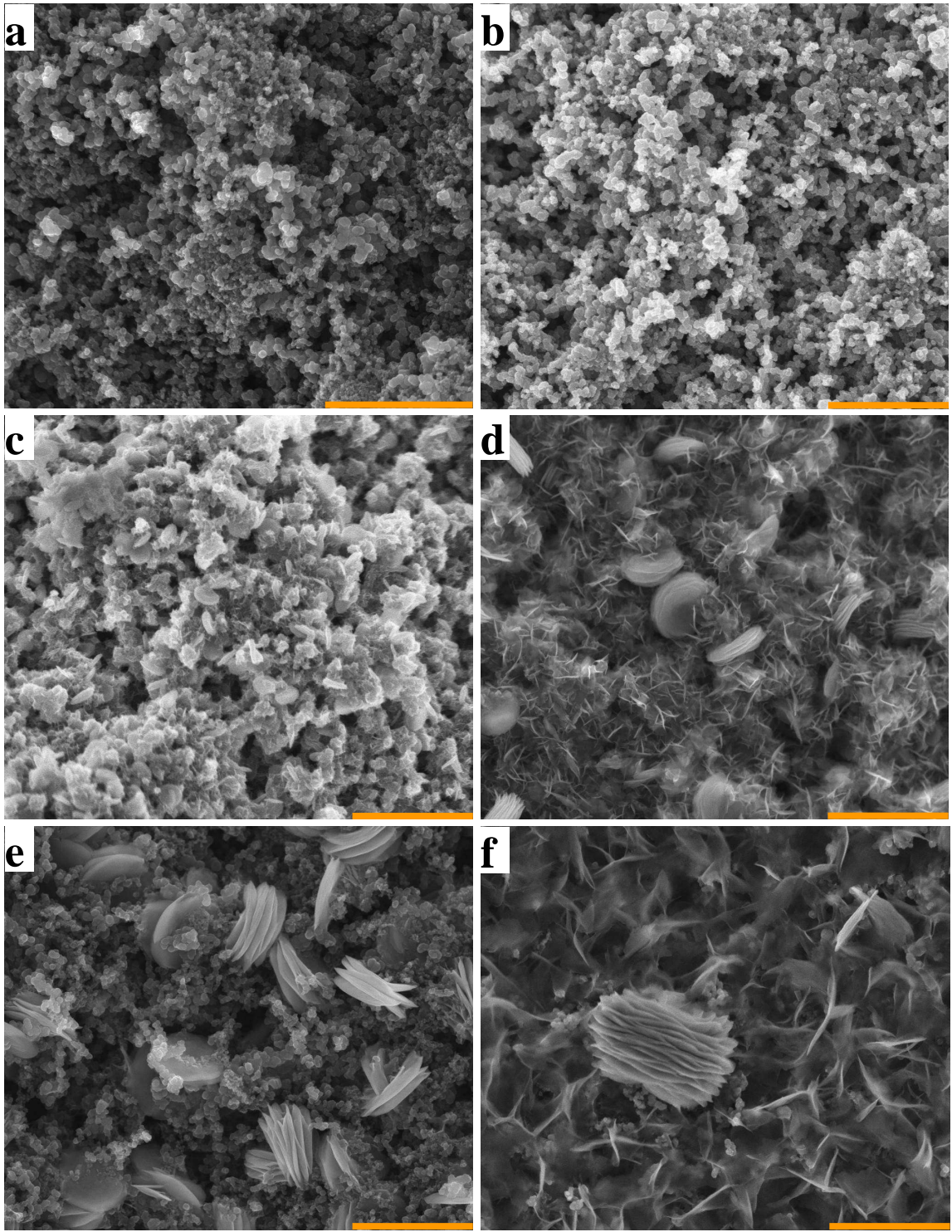

**Figure 1, Aetukuri *et. al*.**

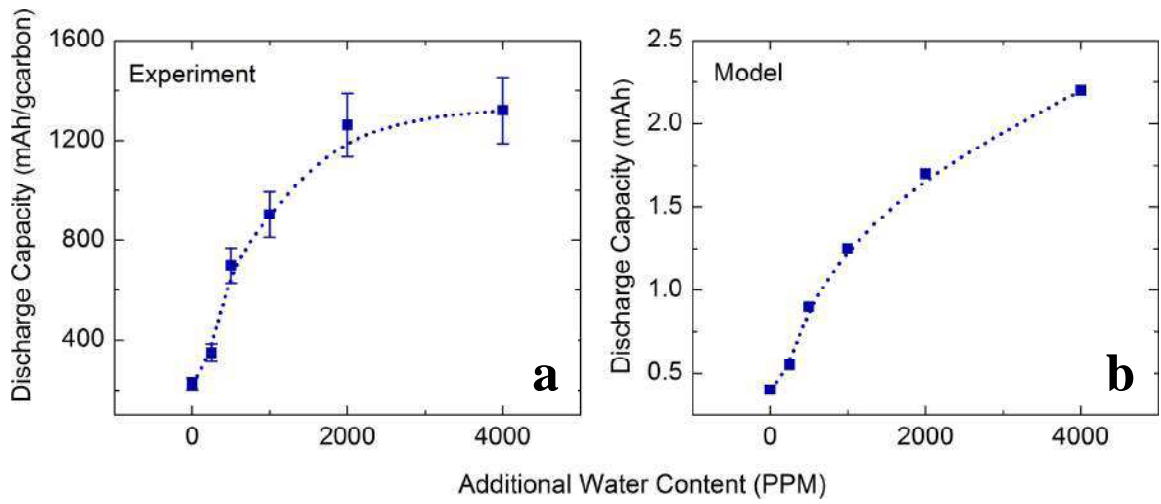

**Figure 2, Aetukuri *et. al*.**

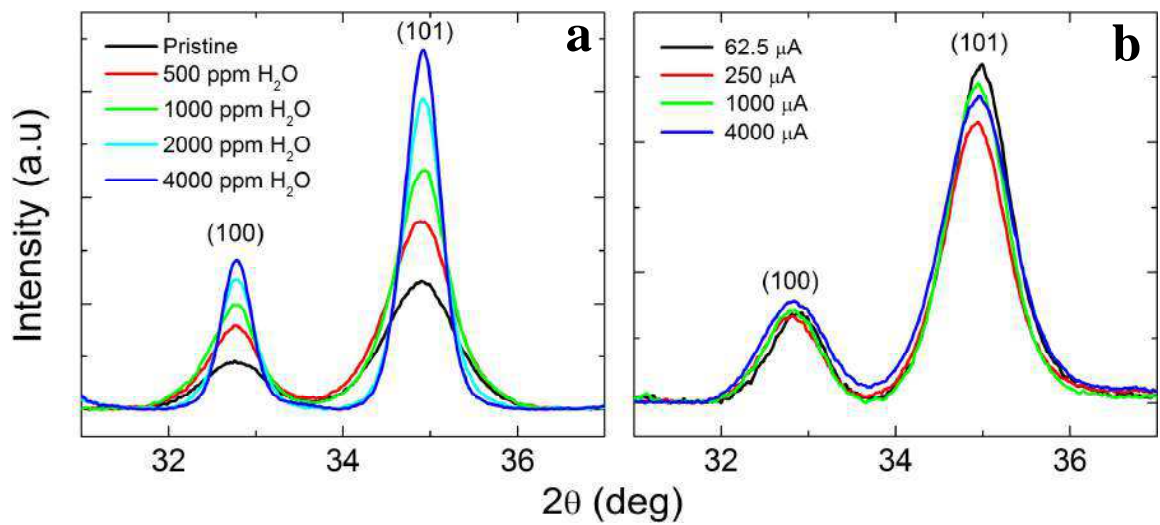

**Figure 3, Aetukuri *et. al*.**

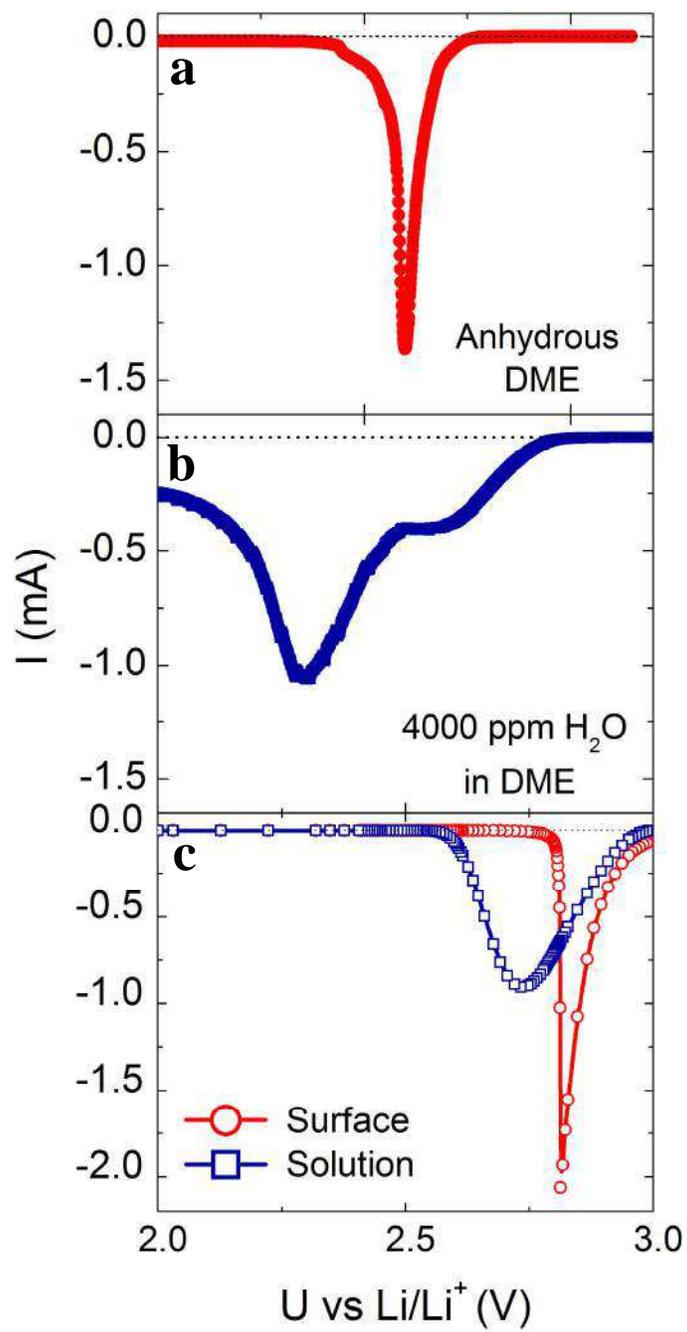

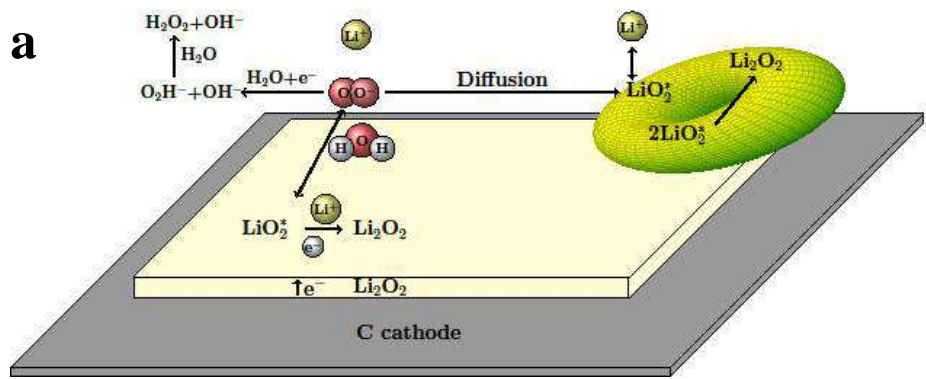

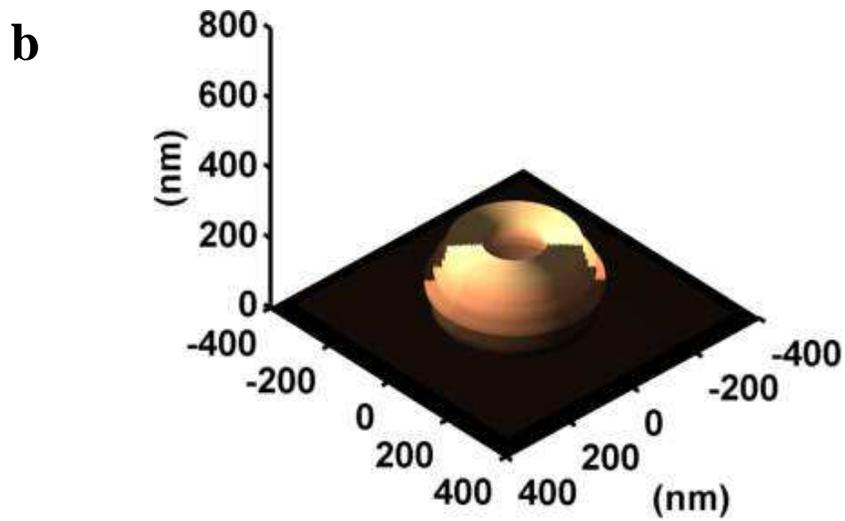

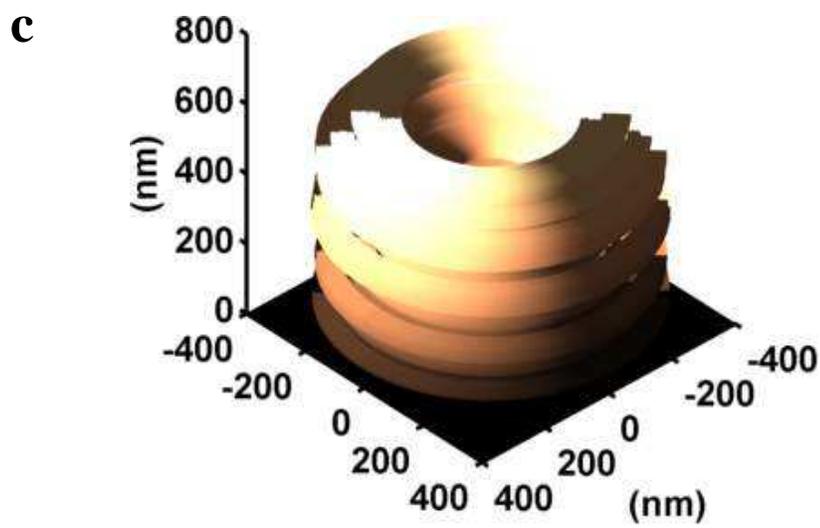

**Figure 5, Aetukuri *et. al*.**

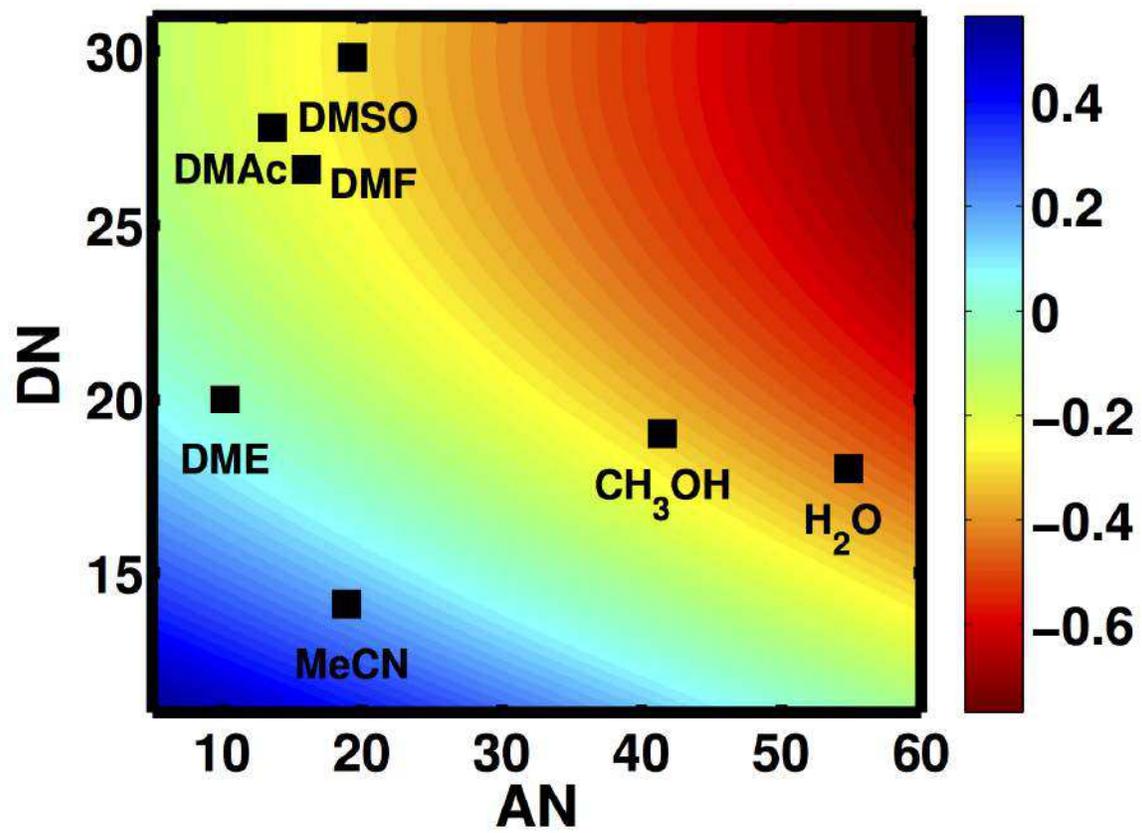

**Figure 6, Aetukuri *et. al*.**

**Figure 1| Li$_2$O$_2$ discharge product morphology control.** Scanning electron microscopy images of **a**, Vulcan XC72® carbon cathode without any discharge and similar cathodes discharged to a capacity of 1 mAh at a rate of 50 µA using **b**, nominally anhydrous (<30 ppm) 1M Li-TFSI in DME as the electrolyte and **c-f**, with water contents of 500 ppm, 1000 ppm, 2000 ppm, 4000 ppm in the electrolyte. The size of the Li$_2$O$_2$ toroids increases with the amount of water in the electrolyte. The thin flake-like features observed in **d** and **f**, we presume increase in size and number of layers to form toroids. All the scale bars are equivalent to 1 µm.

**Figure 2| Discharge capacity increase with increasing water content in the electrolyte. a,** Experimental discharge capacities for batteries employing XC72 carbon cathodes and 1M Li-TFSI in DME with varying water contents as the battery electrolyte. The experimental discharge capacities were obtained from galvanostatic discharges to a reductive potential of 2.3 V (vs Li/Li$^+$) at a discharge rate of 200 µA. b, Theoretically predicted discharge capacities from the developed electrochemical model. A cathode surface area of ~200 cm$^2$ has been assumed for the capacity calculation. The model predicts a ~5 fold enhancement due to the addition of water through the triggering of the solution mechanism. The dotted lines are a guide to the eye.

**Figure 3|** *Ex-situ* **X-ray diffraction measurements on discharged cathodes. a**, θ-2θ x-ray diffractograms near the Li$_2$O$_2$ (100) and (101) resonance peaks on P50 carbon cathodes discharged at a rate of 250 µA in cells employing 1M Li-TFSI in DME with varying water contents (shown in the legend) as electrolytes. The diffractograms show a strong narrowing of the (100) and (101) peaks with increased water content in the electrolyte which is consistent with increasing Li$_2$O$_2$ crystallite size evident from the SEM images of Fig. 1. **b**, Similar diffractograms collected on cathodes discharged at varying discharge rates, as shown in the legend, by employing the nominally anhydrous 1M Li-TFSI in DME electrolyte. No changes in the peak width were observed even by a nearly 2 orders change in the magnitude of the discharge current. This suggests that the Li$_2$O$_2$ remains crystalline but the changes in the crystallite size, if any, are below the instrumental resolution in all samples discharged under anhydrous conditions,

irrespective of the discharge rate. All cathodes were discharged to a discharge capacity of 2 mAh and all the curves are normalized to the carbon cathode's (002) x-ray diffraction peak (not shown).

**Figure 4| The two pathways for $Li_2O_2$ formation.** Discharge linear scanning voltammograms (LSV) performed at 0.05 mV/s with a Vulcan XC72® carbon cathode and lithium anode employing 1M Li-TFSI based electrolyte solutions in **a**, nominally anhydrous DME **b**, DME with 4000 ppm water. The anhydrous DME sample shows a single sharp peak in the LSV at ~2.5 V. By contrast, the LSV curve for the cell with 4000 ppm water exhibits a distinct second peak at ~2.3 V. We attribute the first peak to surface electrochemical growth of $Li_2O_2$ and the second peak to the solution-mediated growth of $Li_2O_2$ where $O_2^-$ acts as a redox mediator. **c,** Theoretically predicted discharge LSV curves for the two independent mechanisms using the developed electrochemical model. The peak currents and the relative potential differences between the two peaks from theory are in good agreement with experiment. The differences in the absolute potentials from theory and experiment could be due to the cell impedance, which is not subtracted from the experimental LSV curves.

**Figure 5| Proposed mechanism for the growth of $Li_2O_2$ toroids in the presence of water. a**, The deposition of $Li_2O_2$ in a Li-$O_2$ cell is shown, schematically, to proceed via a surface electrochemical growth process that occurs on a nucleated film of $Li_2O_2$ through the sequential transfer of $Li^+ + e^-$ to the intermediate species, $LiO_2^*$ and eventually forming $Li_2O_2$. The electron must, therefore, tunnel through the nucleated $Li_2O_2$ film as indicated and this process is limited by the electronic conductivity of $Li_2O_2$. The presence of a solvent that solvates $LiO_2^*$ to $Li^+$ and $O_2^-$ (water in our experiments) triggers a solution pathway leading to the growth of toroids, as shown schematically. The soluble $O_2^-$ adsorbs as $LiO_2^*$ on the growing toroidal particle, ultimately disproportionating to form $Li_2O_2$. Thus, $O_2^-$ acts as a redox shuttle and leads to the formation of large particles thereby circumventing the conductivity limitations in the surface electro-chemical growth. In our experiments, disproportionation of the $O_2^-$ anion in the presence of $H_2O$, a proton source, will also lead to the formation of $H_2O_2$. However, the dominant

electrochemistry is still the formation of $Li_2O_2$ (as substantiated in the main text). $Li_2O_2$ toroidal particle size predicted by the particle growth model developed (and discussed in detail in the SI) for an electrolyte containing **b,** 1000 ppm and **c,** 4000 ppm of water. Larger sized discharge products are observed at higher water contents consistent with the experimental observations.

**Figure 6| Quantitative basis for solvent selection for high capacity Li-$O_2$ batteries.** The free energy of dissolution for $LiO_2^*$ into $Li^+$ and $O_2^-$ in different solvents as a function of the Gutman acceptor and donor numbers (AN and DN). The free energy plot is normalized relative to that of pure DME. Dimethyl formamide (DMF), dimethyl acetamide (DMAc) and dimethyl sulfoxide (DMSO) have high DN and thus are capable of stabilizing $Li^+$. Water and methanol on the other hand, have high acceptor numbers and thus stabilize $O_2^-$. We predict that solvents that fall in the top right quadrant of this plot will favor solution-mediated deposition of $Li_2O_2$, which will be essential for high capacity Li-$O_2$ batteries.

# On the Origin and Implications of Li$_2$O$_2$ Toroid Formation in Nonaqueous Li-O$_2$ Batteries

Nagaphani B Aetukuri[1*], Bryan D McCloskey[1,2,3], Jeannette G Garcìa[1], Leslie E Krupp[1], Venkatasubramanian Viswanathan[4*] and Alan Luntz[1,5]

1 IBM Almaden Research Center, San Jose, CA, 95120
2 Department of Chemical and Biomolecular Engineering, University of California, Berkeley, CA, 94720
3 Environmental Energy Technologies Division, Lawrence Berkeley National Laboratory, Berkeley, CA, 94720
4 Department of Mechanical Engineering, Carnegie Mellon University, Pittsburgh, PA, 15213
5 SUNCAT, SLAC National Accelerator Laboratory, Menlo Park, CA 94025

*Address correspondence to: phani@alumni.stanford.edu (N.B.A); venkvis@cmu.edu (V.V)

# Supplementary Information

**S1. Experimental Methods**

<u>Cathode Preparation</u>

Vulcan® XC72 (from Cabot Corporation) carbon cathodes were prepared by air spraying a slurry made from 1:3 (m/m) 60 wt% PTFE emulsion and XC72 filtered through a 60 mesh sieve in 20:80 (v/v) isopropanol and water mixture onto a stainless steel mesh. Prior to spraying, the stainless steel mesh is cleaned in running water, isopropanol and then acetone and dried at 130 $^o$C under ambient atmosphere. The carbon coated mesh is air dried at room temperature (RT). TiC cathodes were prepared by plastering a slurry comprising of 1:5 (m/m) 60 wt% PTFE emulsion and TiC nanopowder (SkySpring Nanomaterials, ~40 nm) in isopropanol onto a similarly cleaned stainless steel mesh as done for XC72 carbon cathodes. After plastering TiC, the mesh is air dried at RT. 12 mm diameter carbon cathodes were punched out from the air dried stainless steel meshes in the case of XC72 and TiC or from AvCarb® P50 sheet (from the Fuel Cell Store) for P50 cathodes. The punched cathodes are placed in separate glass vials and further rinsed twice in IPA, dried in vacuum at 130 $^o$C for at least 12 hours and transferred to an Ar filled glove box with <0.1 ppm water at all times. In the glove box, the cathodes are rinsed twice in DME and dried at 180 °C for at least another 12 hours before using them in Li-O$_2$ cells. The final carbon loading on the stainless steel mesh is 1.5 – 2 mg.

Electrolytes

Lithium bis(trifluoromethane sulfonyl)imide (Li-TFSI) was the preferred lithium salt for all the experiments reported in this manuscript. 1M Li-TFSI in 1,2 dimethoxy ethane (DME) electrolyte, Li-TFSI salt and the solvents tetraethylene glycol dimethylether (TEGDME) and Dimethyl Sulfoxide (DMSO) were purchased from Novolyte® technologies (now BASF corporation). The water content in all nominally anhydrous electrolytes, measured by Karl-Fischer titration, is <30 ppm. Ultrapure de-ionized (DI) water (18.2 MΩ-cm, Millipore) was used for preparing electrolytes with known quantities of water.

Electrochemical Measurements

All electrochemical measurements were carried out at room temperature on an in-house designed differential electrochemical mass spectrometry (DEMS) system described in detail previously[1-2]. A VMP3 BioLogic multi-channel potentiostat was used for all electrochemical characterization. The DEMS' unique design allows either pressure decay/rise measurements or mass spectrometry to be used to quantify gas consumption and evolution from battery cells. Hermetic electrochemical cells, based loosely on a Swagelok®-type battery cell, are designed to seal by compressing o-rings against a quartz-tube that houses all cell contents. Cells are assembled in an argon glove box with <0.1 ppm water. 11 mm diameter lithium discs punched from an as received 250 μm thick lithium foil (Lectro® Max100 from FMC) were used as anodes for all the cells in this study. Whatman® QM-A grade quartz filters were used as porous electrode-separators when DMSO-based electrolytes were employed and Celgard® 2500 separators were used at all other times. 75 μl of the electrolyte is used when the QM-A separator is used and 65 μl of the electrolyte when the Celgard® separator is used. A 1 mm in height stainless steel ring is used to incorporate a headspace above the cathode. Capillaries soldered into the cathode side of the cell allow gases to be swept through the cell. All discharges (reduction) were done under constant volume conditions in pure oxygen (research purity grade oxygen from Matheson Tri-Gas®) with a starting pressure of ~1200 Torr. Pressure decay was monitored during discharge to calculate oxygen consumption. Battery charge was carried out in ~950 Torr Argon (research purity grade from Matheson Tri-Gas®) under constant pressure conditions after flushing out any residual oxygen in the cell. All gases (Ar and $O_2$) were passed through Pur-Gas in-line moisture traps (gas purity >6.0, Matheson Tri-Gas®) prior to being fed to the DEMS, and only copper or stainless steel tubing was used to feed gases to the cell. In between measurements, all gas lines in the DEMS were kept under vacuum and the system was used only if out-gassing under vacuum was not observed (as monitored using the in-line pressure transducers). All connections in the feed lines, DEMS, and a cell assembled without any battery components were He leak-checked under vacuum to ensure complete hermetic integrity.

Chemical Titrations

The chemical titrations performed for these studies are based on the protocol developed by McCloskey *et al*[1-2]. We only give a brief description of the procedure here. All titrations were performed on XC72 cathodes discharged in DME-based electrolytes with varying water contents as noted in the main text. Within 5 min after a discharge, the cathodes were extracted from the cells in an argon glove box (<0.1 ppm water), transferred into a glass vial and placed in a vacuum chamber, connected to the glove box, for 30 min to evaporate volatile constituents including DME, the electrolyte solvent. Next, the glass vial is sealed with a silicone septa lid and transferred out of the glove box. ~2 mL of ultrapure DI water (18.2 MΩ-cm, Millipore) is then injected into the sealed vial using a syringe. The vial contents are stirred for ~30 seconds to improve the reaction rate of the discharge products with water. The base formed by the reaction of the discharge products with water is titrated using standardized 0.005M HCl solution and phenolphthalein as the end-point indicator.

Hydrogen peroxide ($H_2O_2$) formed by the reaction of $Li_2O_2$ formed during discharge with the water added to the vial is determined by an iodometric titration. For this titration, 1mL of 2 wt% KI in $H_2O$, 1mL of 3.5 M $H_2SO_4$ and 50 µL of a molybdate-based catalyst are added to convert the analyte, $H_2O_2$, to $I_2$. This turns the solution to a pale yellow color. The $I_2$ thus formed is immediately titrated with 0.01N $NaS_2O_3$ until the solution turns a faint straw color. ~0.5 mL of 1% starch solution is now added for precise end-point detection. The solution turns dark blue after the addition of starch and the titration is resumed and continued until it turns clear. A Metrohm Dosino 800 automatic dispenser was used for the peroxide titration, a Hirschman Solarus burette was used for the acid-base titration. All chemicals used for titrations are purchased from Sigma-Aldrich®.

X-ray Diffraction Measurements

All X-ray measurements were performed on P50 cathodes discharged in DME-based electrolytes. P50 was chosen for x-ray measurements as these cathodes allowed us to discharge the cells to higher capacities (ca. 2 mAh) which is essential for high signal to noise ratio. Cells discharged to 2 mAh were transferred to an argon glove box (<0.1 ppm water) and the cathodes were extracted and placed in a glass vial. The cathodes were twice rinsed in DME to wash off the electrolyte salt and the residual solvent was evaporated in a vacuum chamber, connected to the glove box, for ~5 min. The glass vials are moved back to the glove box from the vacuum chamber and the dried cathodes are placed in a custom-built x-ray cell which is o-ring sealed with a kapton® polyimide film. The assembled x-ray cell with the discharged cathode is used for x-ray measurements.

X-ray diffraction (XRD) measurements were performed on a Bruker D8 Discover X-ray diffractometer fitted with a 2-dimensional (2D) X-ray detector. All scans were performed with the detector and incident beam in a symmetric θ-2θ geometry using graphite monochromated Cu-Kα X-rays (λ = 1.5418 Å) collimated in a pin-hole collimator to yield ~650 µm diameter X-ray beam on the sample being measured. Data is collected at room temperature in the 2D mode with an integration time of at least 30 minutes for each frame. During measurements, the discharged cathode is oscillated in the x-y plane (the sample plane) with 2 mm oscillation amplitude. Therefore, the X-ray diffractograms represent a spatial average over an area of at least 4 X 4 mm$^2$ on the cathode. The collected data (at least four frames to cover a 2θ range of 80 degrees) is integrated over χ, the polar angle orthogonal to 2θ to yield the intensity vs 2θ plots shown in the manuscript. We did not find any noticeable changes in the XRD patterns over the measurement time.

Scanning Electron Microscopy

Discharged cathodes used for scanning electron microscopy (SEM) measurements are treated similarly to the cathodes used for x-ray measurements. The dried cathodes are mounted on an SEM sample holder and taken to the SEM sample loading chamber (connected to the SEM) in a sealed glass bottle. The sample holder is then transferred to the sample loading chamber and the latter is then pumped out before transferring the sample holder for imaging in the SEM. The time from opening the glass bottle to the commencement of loading chamber pump down is < 5 seconds. While this procedure prohibits us from performing any quantitative chemical analysis in the SEM, we assume that the change in morphology of the Li$_2$O$_2$ particles is negligible for such a short time exposure to ambient. The measurements were performed on an FEI Helios Nanolab 400s system. Imaging was done at an electron current of 43 pA and an accelerating voltage of either 3 or 5 kV. The images presented in this manuscript were collected with a through-lens detector.

**S2. Supplementary Text and Figures: Experiment**

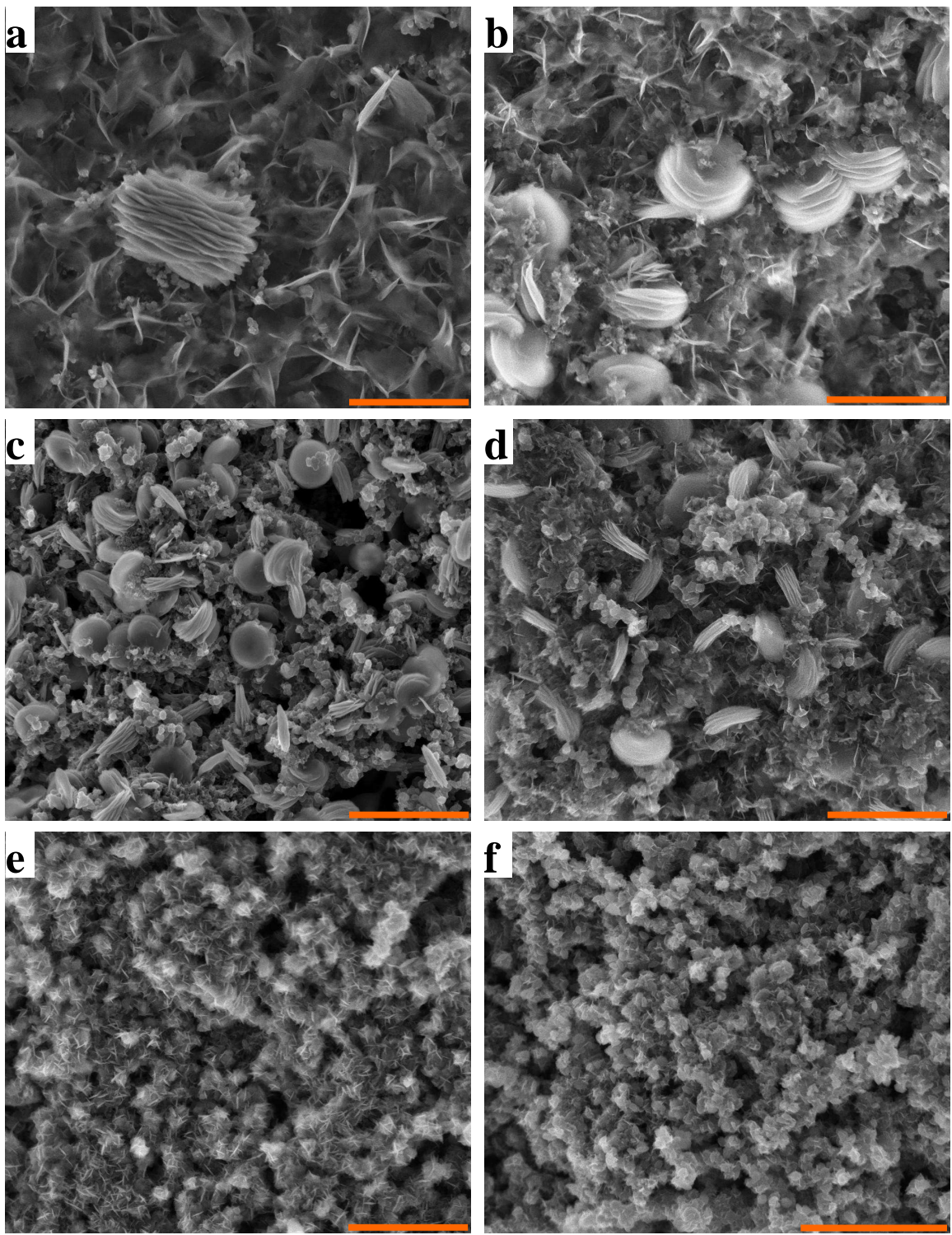

**Figure S1| Li$_2$O$_2$ toroid size as a function of discharge current.** SEM images collected on discharged XC72 cathodes from cells employing 1M Li-TFSI in DME with 4000 ppm

water as an electrolyte and discharged to a capacity of 1 mAh at a discharge rate of **a,** 50 µA **b,** 100 µA **c,** 200 µA **d,** 400 µA **e,** 800 µA and **f,** 1600 µA. All scale bars are equivalent to 1 µm. Clearly, the toroid size is strongly correlated with the discharge rate.

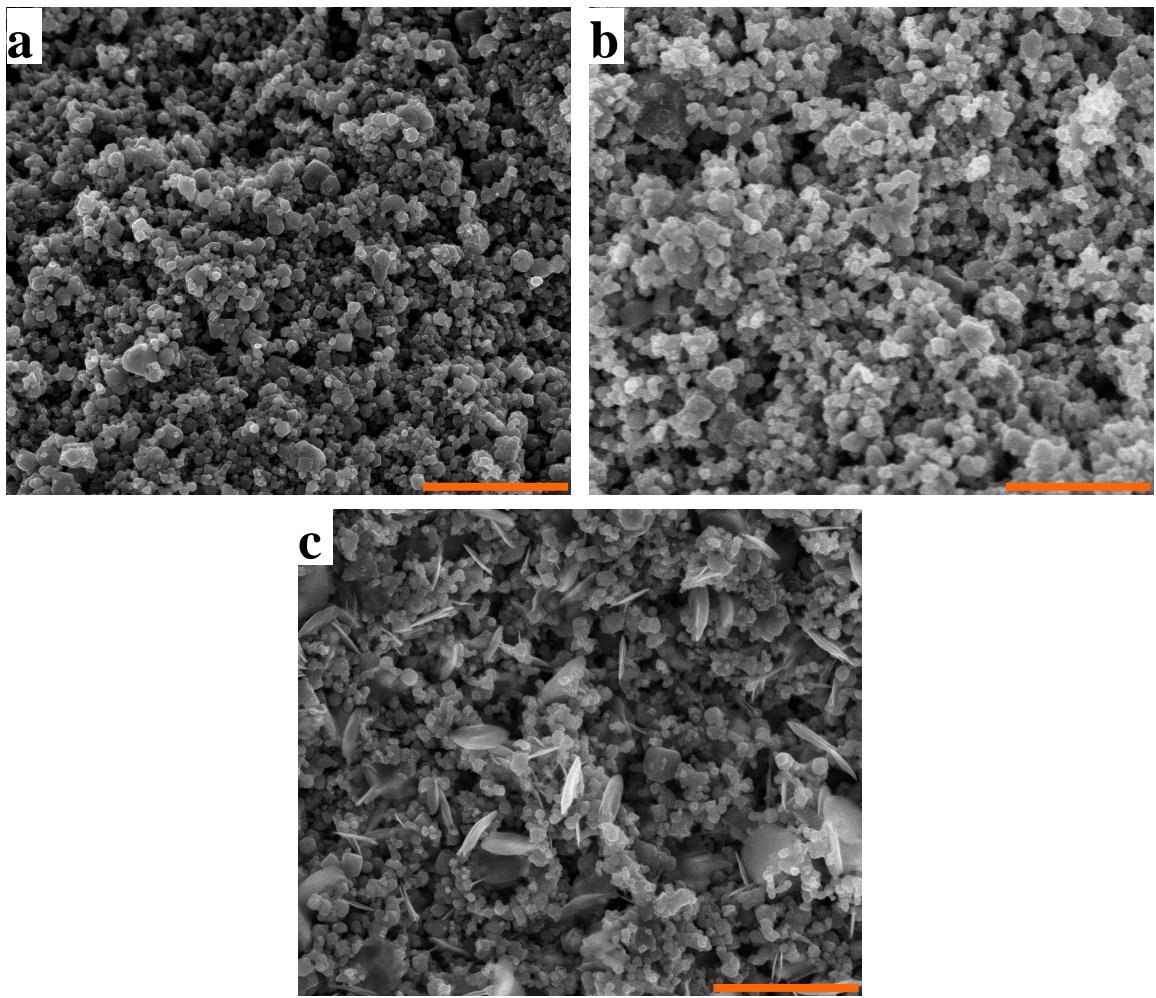

**Figure S2| Li$_2$O$_2$ morphology control on TiC cathodes.** SEM images of **a**, TiC cathode without any discharge and similar cathodes discharged at a rate of 200 µA using **b**, nominally anhydrous (<30 ppm) 1M Li-TFSI in DME as the electrolyte and **c**, with a water content of 4000 ppm in the electrolyte. The size of the Li$_2$O$_2$ toroids directly correlates with the amount of water in the electrolyte showing that the morphology changes are not restricted to a particular cathode material. The cathode in **b**, is discharged to a capacity of 0.83 mAh (full discharge capacity at 200 uA) and in **c**, to 1mAh. All the scale bars are equivalent to 1 µm.

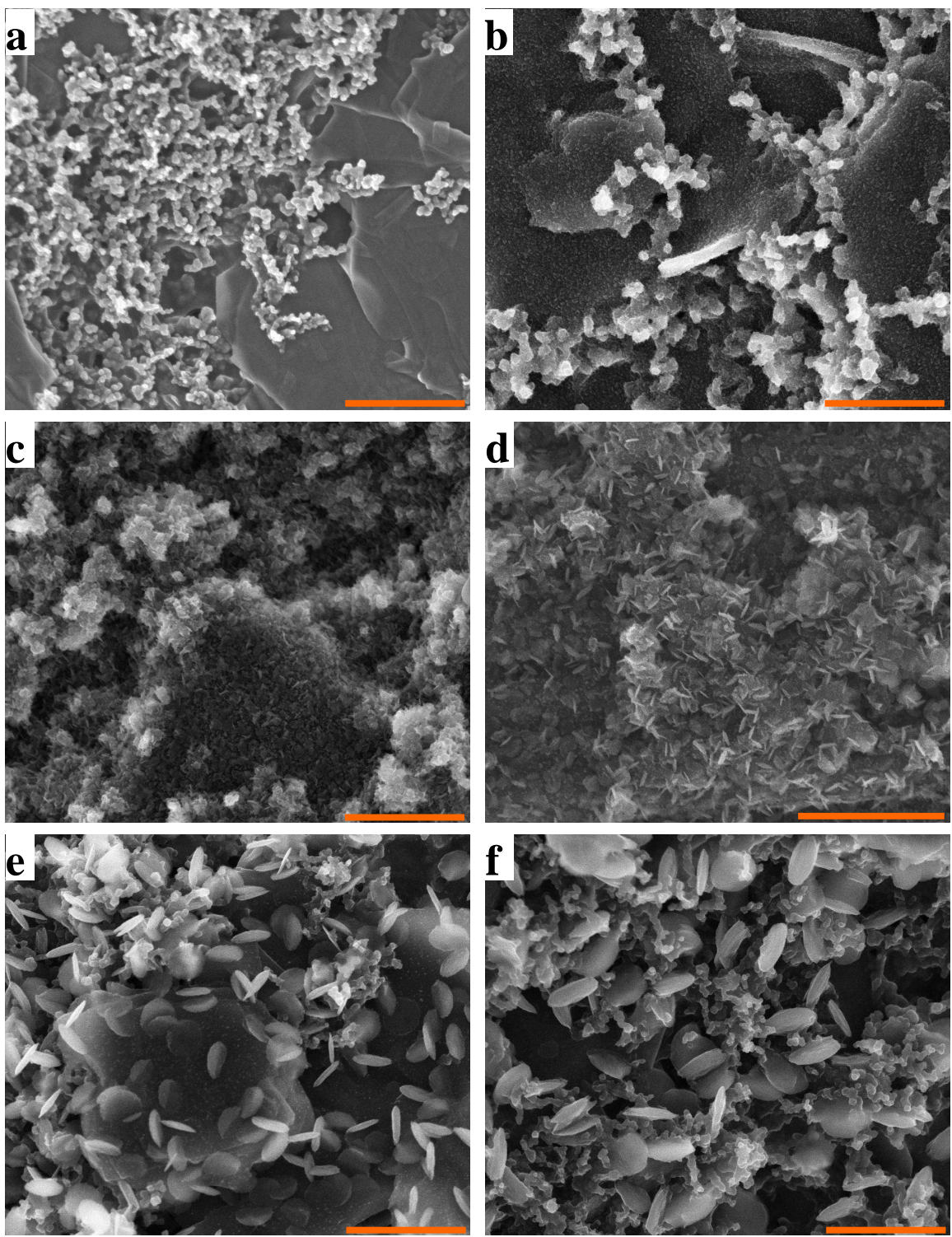

**Figure S3| Li$_2$O$_2$ morphology control on P50 cathodes.** Scanning electron microscopy images of **a**, P50 carbon cathode without any discharge and similar cathodes discharged to a capacity of 2 mAh at a rate of 250 μA using **b**, nominally anhydrous (<30 ppm) 1M

Li-TFSI in DME as the electrolyte and **c-f**, with water contents of 500 ppm, 1000 ppm, 2000 ppm, 4000 ppm in the electrolyte. This toroid-size – water-content correlation observed here is entirely consistent with that presented on XC72 carbon cathodes in the main text (Fig. 1). The SEM imaging on these cathodes was performed after XRD measurements (presented in Fig. 3a of main text) were performed. All the scale bars are equivalent to 1 µm. SEM imaging on the electrode seperators (Celgard and QM-A) did not show any toroids even with the highest water content in the electrolyte.

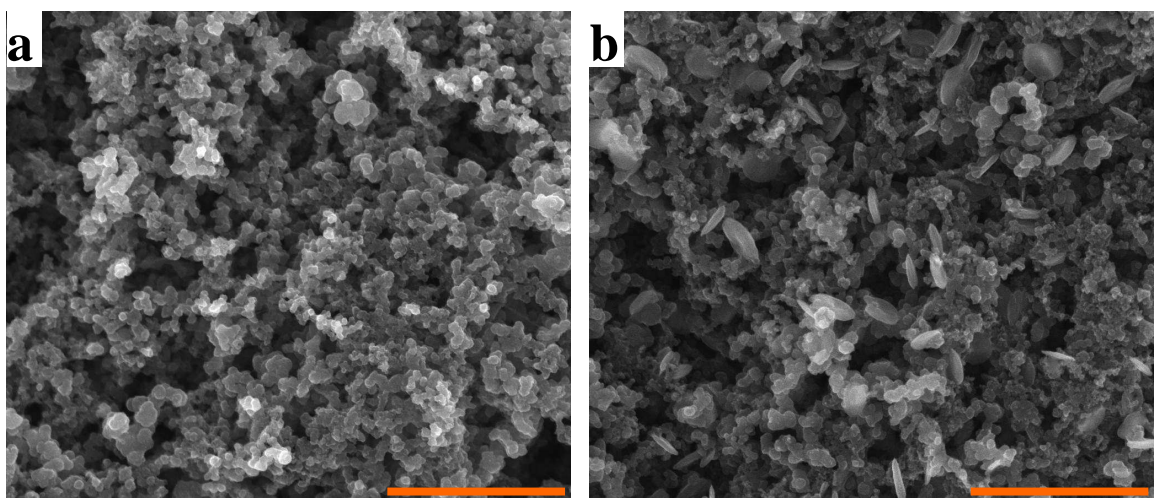

**Figure S4| Li$_2$O$_2$ morphology control with TEGDME as the electrolyte solvent.** SEM images of XC72 cathodes discharged at a rate of 200 µA using **a**, nominally anhydrous (<30 ppm) 1M Li-TFSI in TEGDME as the electrolyte and **b**, with a water content of 4000 ppm in the electrolyte. The size of the Li$_2$O$_2$ toroids directly correlates with the amount of water in the electrolyte showing that the morphology changes are similar to those observed with DME. The cathode in **a** is discharged to a capacity of 0.46 mAh (full discharge capacity at 200 uA) and in **b** to 1mAh. All the scale bars are equivalent to 1 µm.

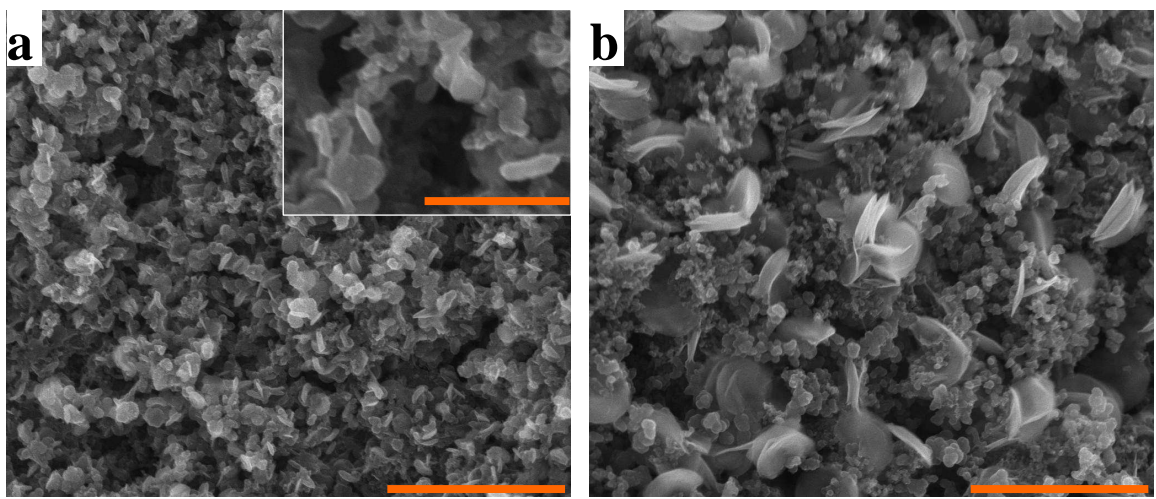

**Figure S5| Li$_2$O$_2$ morphology control with DMSO as the electrolyte solvent.** SEM images of XC72 cathodes discharged at a rate of 200 µA to a capacity of 1 mAh using **a**, nominally anhydrous (<30 ppm) 1M Li-TFSI in DMSO as the electrolyte and **b**, with a water content of 4000 ppm in the electrolyte. The size of the Li$_2$O$_2$ toroids directly correlates with the amount of water in the electrolyte showing that the morphology changes are similar to those observed with DME. However, in the case of DMSO very small toroids, few tens of nanometers in diameter, can be seen (see inset in **a**) even in the nominally anhydrous case. The scale bars in **a** and **b** are equivalent to 1 µm. The scale bar for the inset in **a** is equivalent to 300 nm.

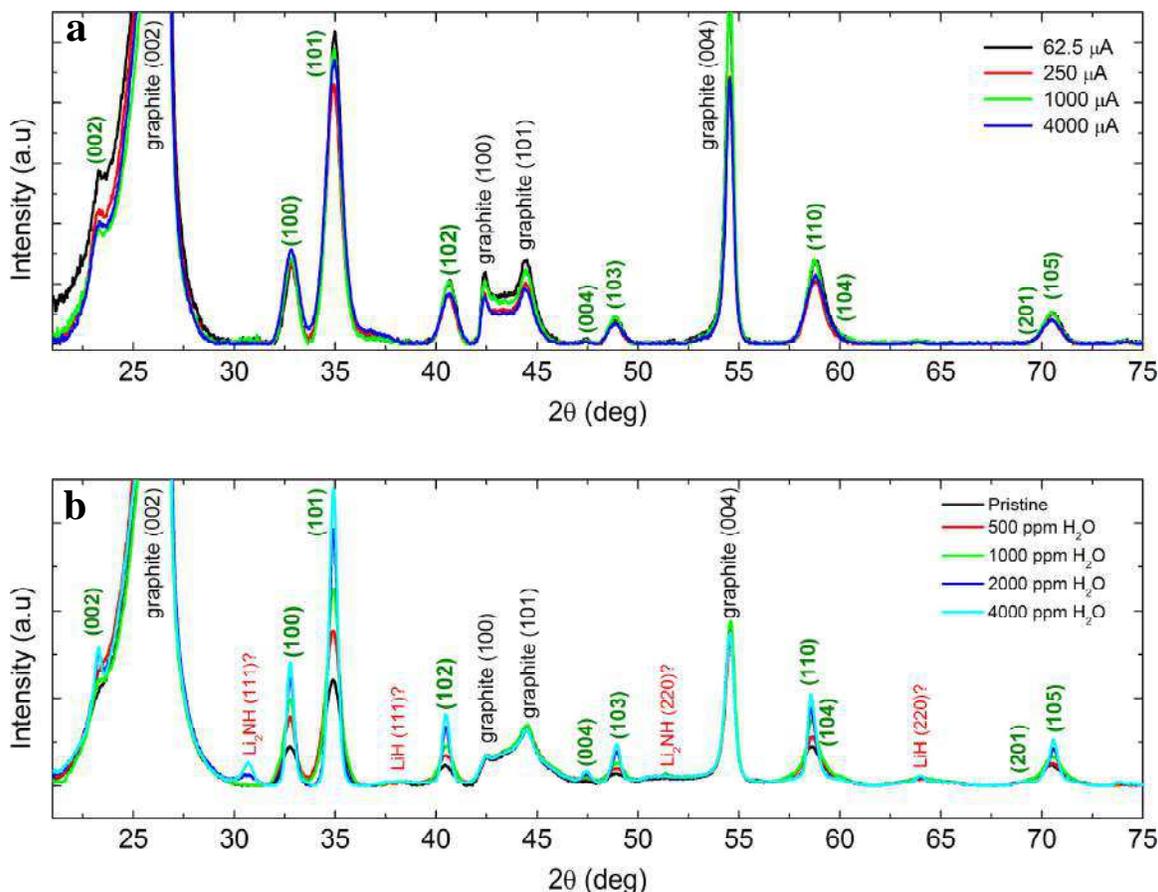

**Figure S6| Discharge rate and water-content dependent crystallinity changes. a**, Wide angle x-ray diffraction θ-2θ plot for cells employing the nominally anhydrous electrolyte (1M Li-TFSI in DME) and discharged at currents from 62.5 μA to 4000 μA, as shown in the legend. All peaks can be indexed[3] to $Li_2O_2$ (labels in green color font) and graphite (from the P50 carbon cathode). **b**, Similar XRD θ-2θ plot for cells discharged at 250 μA with the nominally anhydrous electrolyte and electrolytes with added water from 500 ppm - 4000 ppm, as shown in the legend. In addition to the peaks that can be indexed[4-5] to $Li_2O_2$ (labels in green color font) and graphite, an impurity peak at ~30.65 degrees can be tentatively attributed to be the $Li_2NH$ (111) peak[4-5]. Other possible impurity peaks are also indexed in this diffractogram. All cells were galvanostatically discharged to a discharge capacity of 2 mAh. The slight difference in the graphite (100) and (101) peaks in **a** and **b** are presumably because the P50 carbon cathodes are from two different batches.

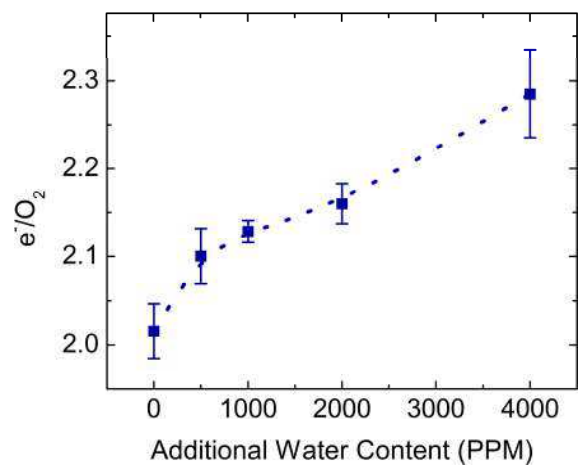

**Figure S7| Discharge electrochemistry in electrolytes with water additive.** A plot of the number of electrons consumed per oxygen molecule, obtained from pressure decay measurements during cell discharge in pure oxygen, as a function of water content in the 1M Li-TFSI in DME electrolyte. The $e^-/O_2$ values are an average over at least five different cells for each water concentration and the error bars represent the standard deviation. The electrochemical formation of $Li_2O_2$ involves two electrons per oxygen molecule. The deviation from this ideal number with increasing water content is indicative of parasitic electrochemistry that is occurring in the cell at high water contents. The dashed line is a guide to the eye.

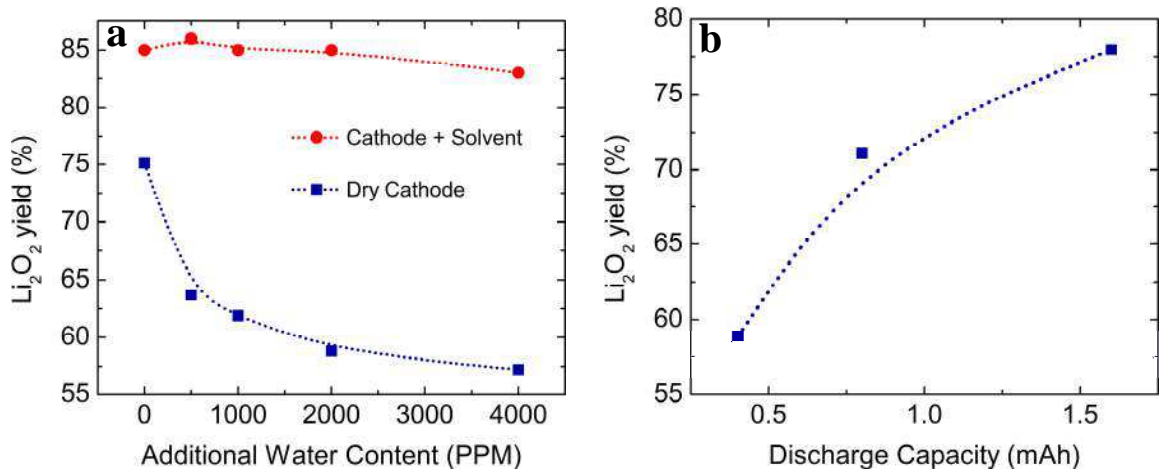

**Figure S8| Summary of chemical titrations. a**, A plot of the $Li_2O_2$ yields obtained from iodometric titration experiments performed on XC72 cathodes discharged to a capacity of 400 μAh at a discharge rate of 100 μA $Li_2O_2$ by employing 1M Li-TFSI in DME based electrolytes with varying water contents as shown on x-axis. The cathodes were dried in vacuum for 30 min (see methods section for details). $Li_2O_2$ yield is defined as the percentage of the titrated peroxide quantity to the expected quantity of $Li_2O_2$ based on the discharge capacity. Also, included are the titrations performed on the cathodes with most of the electrolyte solvent not evaporated. We take the difference in the titrated peroxide quantities under dry and wet (cathode+solvent) conditions as an indirect measurement for the amount of $H_2O_2$ generated by the disproportionation of $O_2^-$ in the presence of water. **b**, The titrated $Li_2O_2$ quantity as a function of discharge capacity for cells employing 1M Li-TFSI in DME with 2000 ppm added water as an electrolyte. This data suggests that most of the capacity enhancement in the batteries with added water is due to the formation of $Li_2O_2$. Note that the $Li_2O_2$ yield is a function of the discharge rate even for the anhydrous sample. For example, samples discharged to 400 uAh at a discharge rate of 100 uA and 500 uA showed $Y_{Li2O2}$ values of 75% and 82% respectively. This is consistent with our previously published results[6]. The dotted lines in the both the plots are a guide to the eye.

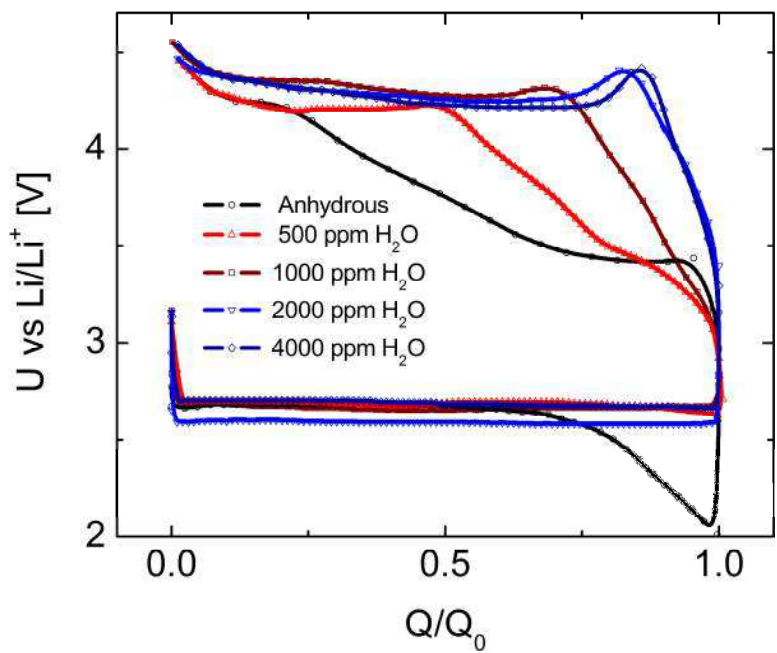

**Figure S9| Charge potential variation with water content in the electrolyte.** Discharge-charge curves for cells with XC72 cathodes that are discharged at a rate of 200 uA and employing 1M Li-TFSI in DME with varying water contents (shown in the legend) as the electrolyte. The x-axis is normalized to the full discharge capacity ($Q_0$) of the cathodes, which is 0.433 mAh for the cell employing a nominally anhydrous 1M Li-TFSI in DME as the electrolyte and 1 mAh for all the others. Clearly, the charge over potential increases rapidly to a value >4.2 V in the cells with water contaminated electrolytes. We believe that the charge over-potential is mainly due to the parasitic discharge products whose formation is accelerated in the presence of water as shown in Fig. S6 and S7.

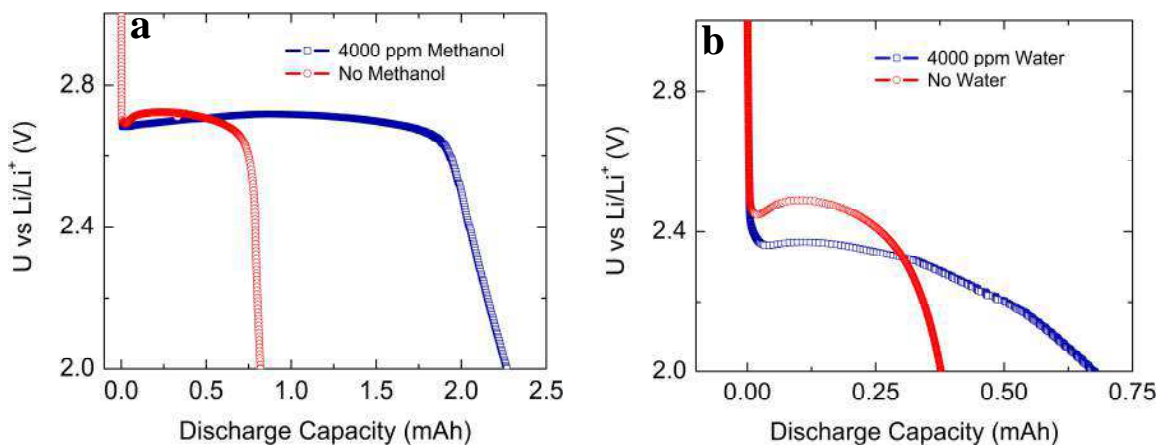

**Figure S10| Discharge capacity comparison for methanol and water as additives in DME. a,** Galvanostatic discharge curves at a discharge rate of 100 µA for cells employing 1M Li-TFSI in anhydrous DME and DME with 4000 ppm methanol as electrolytes. The total discharge capacity increases by 3 times in the presence of Methanol. In Fig. 6, we predicted that methanol, like water, has a high acceptor number and will therefore increase the solubility of $O_2^-$. This additional route to the formation of $Li_2O_2$, we suggest, is the reason for enhanced capacity in the presence of methanol. **b,** Similar galvanostatic discharge curves at a discharge rate of 3 mA for cells employing 1M Li-TFSI in anhydrous DME and DME with 4000 ppm water as electrolytes. Even at this high discharge current of 3 mA, the increase in capacity due to water addition is close to 2 times the capacity obtained in the anhydrous case. Notably, with methanol as an additive and for water at high currents, we did not observe any $Li_2O_2$ toroid formation. Therefore, we suggest that the benefits of capacity enhancement by solution-mediated deposition of $Li_2O_2$, are not restricted to low discharge rates.

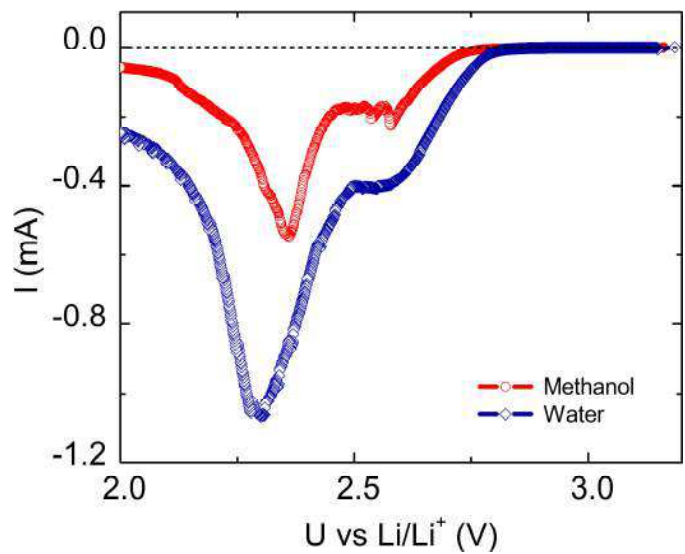

**Figure S11| Linear scanning voltammograms with water and methanol.** Discharge linear scanning voltammograms performed at 0.05 mV/s with a Vulcan XC72® carbon cathode and lithium anode employing 1M Li-TFSI based electrolyte solutions with 4000 ppm methanol (red) and 4000 ppm water (blue) as additives. The two LSV curves exhibit a distinct second peak at potentials <2.5 V, although with differing intensities. These LSV curves are consistent with our theoretical modeling based on the hypothesis that solvents with high acceptor number activate the solution-mediated growth of $Li_2O_2$ where $O_2^-$ acts as a redox mediator.

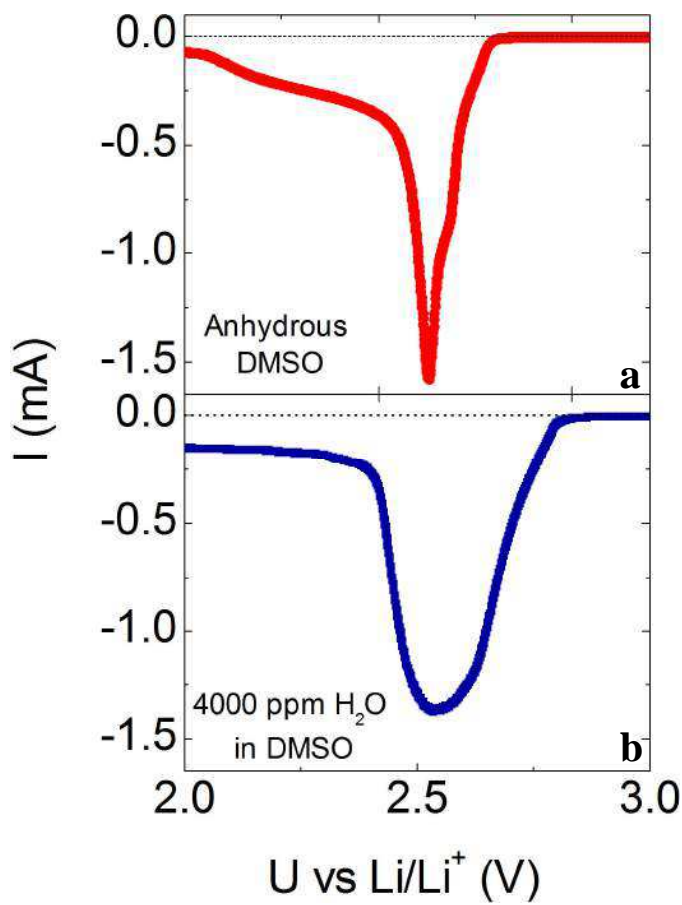

**Figure S12| Discharge LSV curves for DMSO.** The results of discharge LSV experiments performed at 0.05 mV/s with a Vulcan XC72® carbon cathode and lithium anode employing 1M Li-TFSI based electrolyte solutions in **a**, nominally anhydrous DMSO and **b**, DMSO with 4000 ppm water as the solvents. The anhydrous DMSO sample shows a sharp peak in the LSV at ~2.5 V with a very small, but distinct shoulder at ~2.3 V. As discussed in the main text, we attribute the peak at ~2.5 V to the surface electrochemical growth of $Li_2O_2$ and the one at ~2.3 V to the solution-mediated growth. The presence of this small shoulder correlates with the presence of toroids on the cathode (see Fig. S5a). By contrast, the LSV curve for the cell with 4000 ppm water exhibits a broad peak which is possibly due to the two above mentioned mechanisms of $Li_2O_2$ growth.

## S3. Supplementary Text and Figures: Theory

Battery discharge model

The model for simulating battery galvanostatic discharge is formulated as a set of differential algebraic equations (DAE). The galvanostatic condition is imposed as an algebraic equation defining the reaction rates and the material balances for the species associated with the electrochemical reactions are solved as coupled differential equations.

*Surface electrochemical growth*

The surface electrochemical growth of $Li_2O_2$ follows the sequential transfer of $Li^+$ and $e^-$ as discussed in detail in our earlier work.[3] The mechanism is given by:

$$O_2(g) + Li^+ + e^- \rightleftarrows LiO_2^*$$
$$LiO_2^* + Li^+ + e^- \rightleftarrows Li_2O_2(s)$$

Alternately, the second step could be,

$$2LiO_2^* \rightleftarrows Li_2O_2(s) + O_2(g)$$

This surface electrochemical growth could occur along different kind of sites such as kinks, steps and terraces. The overpotentials associated with the growth on these different sites have been calculated using density functional theory calculations in our earlier work[7] and the same formalism is used to describe the surface electrochemical growth in our model. At the low current density conditions being simulated here, the surface growth of $Li_2O_2$ proceeds primarily along the kink and step sites of $\eta = 0.19$ V and $\eta = 0.28$ V respectively[8] on the dominant (0001) surface (that is half $O_2$ converted at equilibrium). We assume the kinetic current density is described by the Tafel equation and is given by:

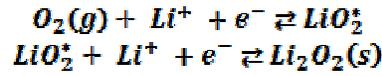

$$i_{Li_2O_2} = i_0 e^{-\max\left(\frac{(U-U_i)}{b_1}, 0\right)}$$

where,

$$U_k = U_0 - \eta = 2.85 - 0.19 = 2.66V \text{ and,}$$
$$U_s = U_0 - \eta = 2.85 - 0.28 = 2.57V$$

respectively, for the growth along the kink and step sites and $b_1$ refers to the Tafel slope, here chosen to be 120 mv per decade. When the potential, U is raised below $U_i$, the reaction becomes exergonic and loses its potential dependence and the current is given by the prefactor, $i_0$. It must be noted that the ratios of kink and step sites need to be included to determine the overall rate and this is given by

$$i_{Li_2O_2,SURF} = \sum_k \theta_k i_{Li2O2,k}$$

where the index, k ∈ {kink, step, terrace} and $\theta_k$ represents the surface coverage of site $k^7$. It is essential to note that the surface electrochemical growth includes the rate of electron transport through the discharge product, $Li_2O_2$. In our earlier work, in the tunneling dominated regime, the rate of electron transport decays exponentially with the thickness of the discharge product[8]. The tunneling limit imposes that surface electrochemical growth shuts off at ~10 nm sized discharge products.

*Solution growth*

In this model, we consider a second possible route for the growth of $Li_2O_2$. In water-contaminated cells (or a few other electrolytes/additives), we assume the generation of soluble reduced oxygen species. The mechanism for solution growth is initiated by the generation of soluble intermediate species in the presence of water, given by,

$$LiO_2^* \rightleftarrows Li^+{}_{(aq)} + O_2^-{}_{(aq)}$$

The free energy change associated with this dissolution process is given by,

$$\Delta G_{soln} = G_{LiO_2^*} - G_{Li^+} - G_{O_2^-} = G_{LiO_2^*} + [(G]_{Li} - G_{Li^+}) + [(G]_{O_2(g)} - G_{O_2^-}) - [(G]_{Li} + G_{O_2(g)})$$

In different solvents, it has been shown that the redox potential shift of $Li/Li^+$ scales with the Gutman donor number of the solvent[9]

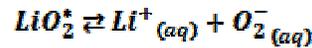
$$[(G]_{Li} - G_{Li^+}) \sim aDN$$

as given in Fig. S13. The stabilization of Li/Li$^+$ saturates beyond a Gutman donor number of ~25 . For the redox potential shift of $O_2/O_2^-$, a similar scaling is observed with the solvent's ability to accept electrons given by the Gutman acceptor number,

$$[(G]_{O_2(g)} - G_{O_2^-}) \sim bAN$$

(see fig. S14). Thus qualitatively, the free energy change associated with the activation of the solution process is given by

$$\Delta G_{soln} \sim aDN + bAN$$

where a ~ 0.1 and b ~ 0.01.

The solution soluble $O_2^-(aq)$ undergoes subsequent reaction on a growing Li$_2$O$_2$ particle leading to the formation of Li$_2$O$_2$ through the reaction given below:

$$2Li^+_{(aq)} + 2O_2^-_{(aq)} \rightleftarrows Li_2O_2(s) + O_2(g)$$

The solution route to the formation of Li$_2$O$_2$ essentially uses the soluble reduced oxygen as a redox shuttle to close the electrochemical cycle. Thus, the rate of electron transport through the discharge product does not directly affect the solution electrochemical growth rate as long as the Li$_2$O$_2$ formed by the surface electrochemical growth remains thin enough such that sufficient current to support the regeneration of O$_2^-$ necessary for the solution electrochemical growth can be drawn through this insulating layer. The solution electrochemical growth depends on the water content as this determines the rate of soluble reduced oxygen species. The rate of solution electrochemical growth is then given by

$$i_{Li_2O_2,SOLN} = i_S c_{O_2^-}^2$$

where $c_{O_2^-}$ is the concentration of solution generated $O_2^-$ and $i_S$ is the prefactor which includes the diffusion rate of $O_2^-$. $i_s$ here is chosen to be 1 µA/cm$^2$.

In our model, we include an additional parasitic electrochemical reaction of the soluble reduced oxygen species with water leading to hydrogen peroxide, given by:

$$O_2^-_{(aq)} + H_2O + e^- \rightleftarrows O_2H^- + OH^-$$

The current associated with this decomposition reaction is given by

$$i_{H_2O_2} = i_{0.H2O2} c_{O_2^-} c_{H2O} e^{-\max\left(\frac{(U-U_{H2O2})}{b_1}, 0\right)}$$

where, $U_{H2O2}$ is the equilibrium potential for the formation of $H_2O_2$. All Tafel slopes are assumed to be 120 mV/dec. Note that the formation of $H_2O_2$ is *not* associated with $Li_2O_2$ formation. It is simply a parasitic electrochemical/chemical reaction. $U_{H2O2}$ is assumed to be 2.6 V and $i_{0,H2O2} = 3$ mA/cm$^2$.

The overall algebraic condition during discharge is given by

$$i = i_{Li_2O_2,SURF} + i_{Li_2O_2,SOLN} + i_{H2O2}$$

where *i* is the galvanostatic current of the cell.

In order to make quantitative comparisons of the results to the model to the experimental results, we also include some consumption of water at the lithium anode (as evidenced by some $H_2$ evolution during open circuit conditions with added $H_2O$). This consumption is assumed to be a chemical reaction obeying first order kinetics in the water concentration.

The algebraic condition imposes material balances for $Li_2O_2$ given by:

$$\frac{dV_{Li_2O_2}}{dt} = \frac{MW_{Li_2O_2} * i_{Li_2O_2}}{F \rho_{Li_2O_2}}$$

where $MW_{Li_2O_2} = 45.881 \frac{g}{mol}$ and $\rho_{Li_2O_2} = 2.31$ g/cc respectively. The other symbols have their usual meaning.

The concentrations of different species are given by the following differential equations:

$$\frac{dc_{O_2^-}}{dt} = k_1 c_{H_2O} - k_{-1} c_{O_2^-} - \frac{(2 i_{Li_2O_2,SOLN} + i_{H2O2})}{F}$$

where $k_1 = 2 \times 10^{-5}$ and $k_{-1} = 10^{-5}$.

$$\frac{dc_{H_2O}}{dt} = -\frac{i_{H2O2}}{F} - k_{Li} c_{H_2O}$$

$$\frac{dc_{HO_2^-}}{dt} = \frac{i_{H2O2}}{F}$$

where $k_{Li} = 10^{-5}$.

Li$_2$O$_2$ Growth Model

The growth model describes the shape of the discharge product, incorporating surface and solution growth. The surface growth is homogenous and shuts down beyond the tunneling thickness. Thus, the surface growth produces a nearly uniform conformal growth on the surface.

Solution growth requires growth on existing Li$_2$O$_2$ site (likely a kink site on an already nucleated film). This relies on the fact that growth of Li$_2$O$_2$ on top of Li$_2$O$_2$ is easier than on C at low currents because of its lower overpotential. In our model, there are three processes that occur simultaneously.

(i) There is a growth rate, $r_{growth}$ associated with the solution process determined by the concentration of soluble reduced oxygen species and its diffusion as discussed above. The overall rate is ultimately set by the current density.

(ii) A parasitic process that leads to the formation of an undesired insoluble product, such as LiOH or Li$_2$NH. This process leads to a surface layer that prevents further growth of Li$_2$O$_2$ determined by the parasitic rate, $r_{par}$.

(iii) A vacancy defect generation rate, $r_{def}$ on the parasitic film on Li$_2$O$_2$, which leads to the nucleation of a new layer when the surface is fully coated by the insoluble parasitic product.

In our model, we assume the defect generation rate is much smaller than the parasitic rate and the growth rate. Thus, this implies a time scale separation and this implies that this process can be simulated sequentially. Thus, the simulation is carried out in the following way.

(i) Initially, a conformal film grows based on surface electrochemical route.

(ii) Subsequently, we pick a defect nucleation site, at a given $r$, chosen by the appropriate probability distribution function (PDF) discussed below.

(iii) We allow for the growth of Li$_2$O$_2$ discharge product both inward and outward from the nucleation site given by the growth rate, $r_{growth}$. At that same time, the surface of the Li$_2$O$_2$ is being covered by the parasitic discharge product given by the rate, $r_{par}$.

(iv) These two processes continue until a time t, when the entire surface is covered by the parasitic product. Growth can only proceed beyond this when another defect site is generated in the parasitic covering film. This is determined by the defect generation rate, $r_{def}$. Using our assumption, that the defect generation rate is much smaller than the growth and parasitic rates, the growth of each separate layer can be time separated and simulated separately. Thus, the simulation reverts back to step 2. This is repeated until the overall capacity (volume) obtained from the detailed electrochemical model is satisfied.

The discussions below use cylindrical symmetry to describe the different rates. We assume that the probability for growth is uniform over the entire surface and for a cylindrical disc, the probability of growth at given a surface element, dr, is given by 2π$r$d$r$, where r is the radius of surface element from the center of some assumed growing particle. Thus, there is a greater likelihood of growth at the edges. Once, nucleated the particle can grow both inward and outward. After a time t, the total inward growth radius, $r_{in}$, and outward growth radius, $r_{out}$, is given by the material balance equation.

$$r_{growth} = \frac{MW_{Li_2O_2} * i_{Li_2O_2}}{F\rho_{Li_2O_2}}$$

$$V = r_{growth}t = \pi(r_n^2 - r_{in}^2) * d_{layer} + \pi(r_{out}^2 - r_n^2) * d_{layer},$$

where V is the total volume of the Li$_2$O$_2$ discharge product grown in time t past nucleation, $r_n$ is the radial location of the nucleation point and $d_{layer}$ is the thickness of the layer. Assuming equal likelihood to grow outward and inward, we get that $r_n$-$r_{in}$ > $r_{out}$-$r_n$. This simply says that per unit of growth, outward growth generates greater surface area. This leads to expressions for both $r_{in}$ and $r_{out}$ as a function of time t.

At the same time, there is a parasitic chemical reaction rate due to the presence of the water. This leads to the covering of the Li$_2$O$_2$ surface and ultimately shutting off the Li$_2$O$_2$ growth process. This parasitic rate depends on the water concentration and is given by:

$$r_{par} = k_{par} c_{H_2O},$$

where $k_{par}$ is the parasitic rate constant. The individual thickness of these layers, $d_{layer}$, is ultimately decided by the interplay between the growth rate and the parasitic rate. At low current rates and high $H_2O$ concentrations, large diameter layers with a small $d_{layer}$ are formed and the layers can then splay apart. However, at higher currents and/or smaller $H_2O$ concentrations, the layers are smaller in diameter, i.e. the tendency to form toroids is suppressed (see Fig S18). This is observed in our simulations and consistent with the experimental results.

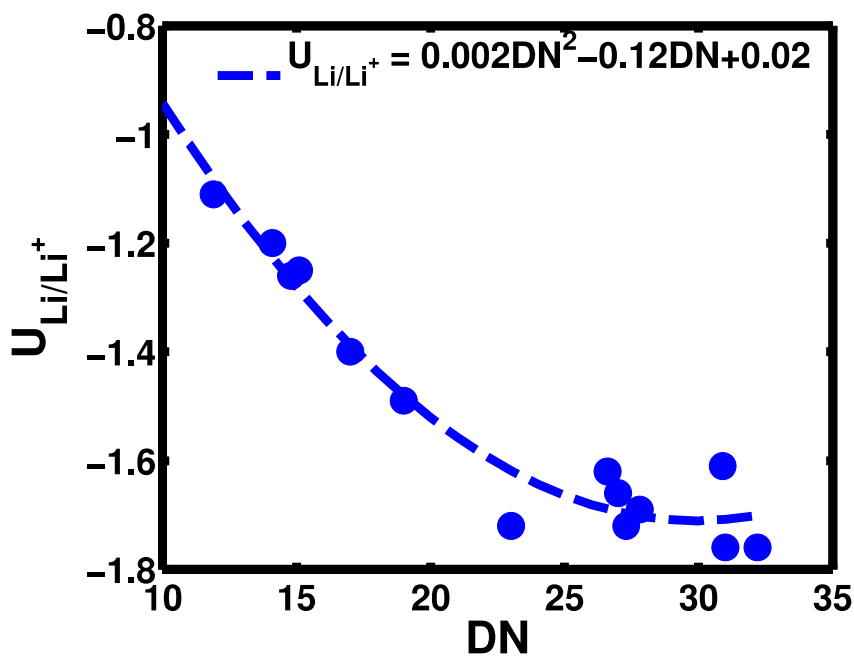

**Figure S13| Li/Li$^+$ redox potential versus Gutman donor number.** A plot of the experimentally measured (filled blue circles) half-wave potentials of the Li/Li$^+$ redox couple in different solvents plotted as a function of the Gutman donor number (DN). The experimental measurements are taken from the half-wave potentials measured by Gritzner et al[10]. The redox potential for the Li/Li$^+$ redox couple is reported relative to Bis(biphenyl)chromium(I)/(0) couple. We observe a nearly linear dependence of the half-wave potential of Li/Li$^+$ in the DN range 10-25. Beyond DN of 25, the theory predicts a saturation in the Li$^+$ solvation leading to a nearly constant value of Li/Li$^+$. A

best fit (quadratic) curve is shown in the plot and this quadratic relation is used in the generalization analysis for solution solubility presented in Fig 6.

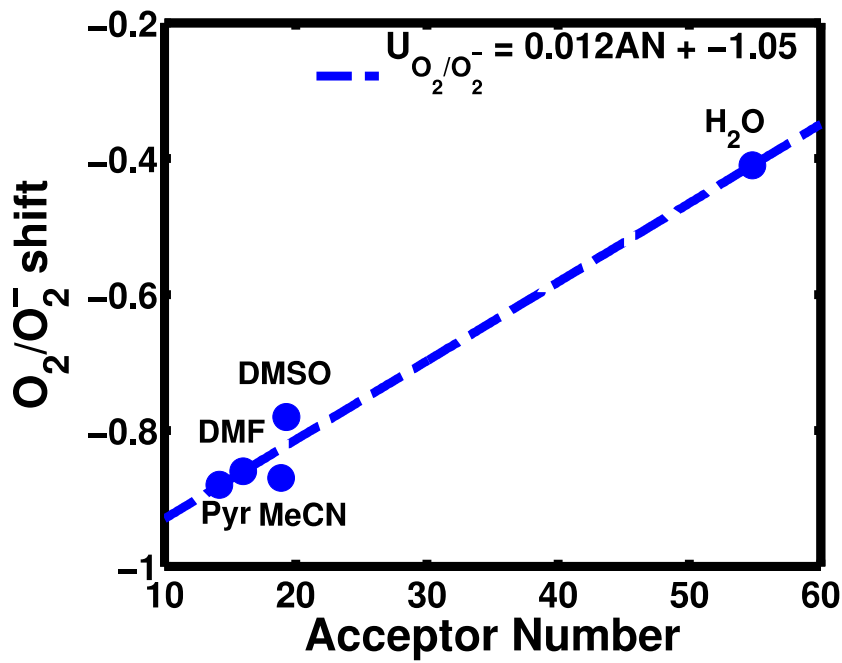

**Figure S14| $O_2/O_2^-$ redox potential versus Gutman acceptor number.** A plot of the redox potential of $O_2/O_2^-$ in different solvents as a function of Gutman Acceptor Number (AN). The experimental measurements for the redox potential have been taken from the work of Sawyer *et. al*[10] The redox potentials of $O_2/O_2^-$ is reported relative to Saturated Calomel Electrode (SCE). We observe a linear dependence of the redox potential with the Acceptor number. It is also to be noted that no trend is observed with DN as is to be expected. The best fit (linear) curve is shown in the plot and this linear relationship is used for the generalization of the solution mechanism discussed in the main text and presented in Fig 6.

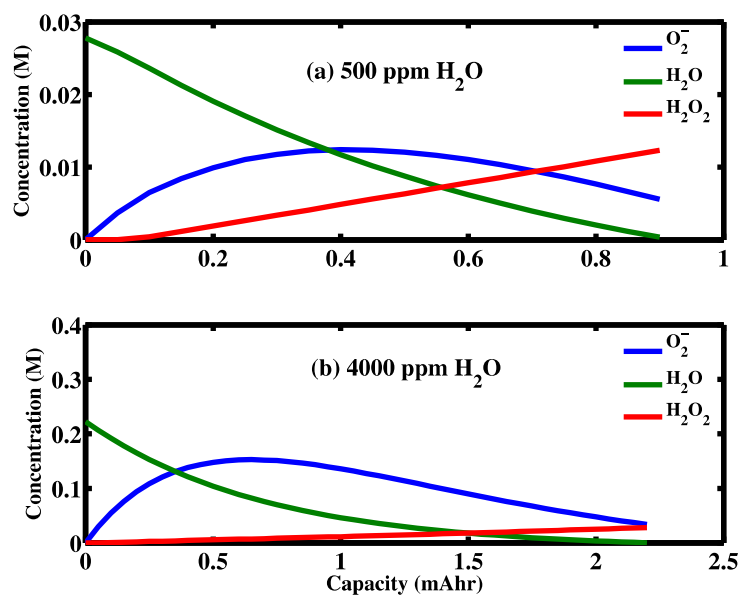

**Figure S15| Concentration changes during discharge.** Concentration profiles of $O_2^-$, $H_2O$ and $H_2O_2$ as a function of capacity for **a,** 500 ppm and **b,** 4000 ppm water in the electrolyte which are obtained from the model under galvanostatic discharge at I = 1 µA/cm². In both cases, initially, there is a surge in the soluble $O_2^-$ concentration. However, as time proceeds, water is being consumed through a parasitic reaction with the lithium metal and through superoxide anion's disproportionation[11] that we account for in the model. The latter leads to the formation of $H_2O_2$, the time-dependent concentration of which is also plotted in the figure. The net decrease in water concentration leads to an overall decrease in the production rate of soluble $O_2^-$ while the galvanostatic condition imposes a nearly constant consumption rate of $O_2^-$. Thus, we observe a net decay in the concentration of $O_2^-$, ultimately leading to cell death.

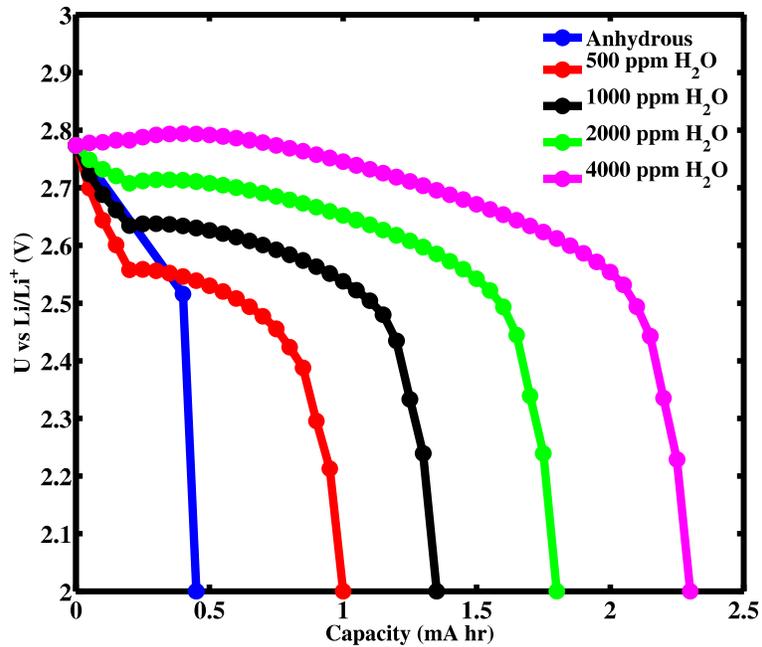

**Figure S16| Theoretically predicted galvanostatic discharge curves**. A plot of the cell potential (U vs Li/Li$^+$) as a function of capacity at three different water concentrations, anhydrous (blue), 500 ppm, 1000 ppm, 2000 ppm and 4000 ppm for a galvanostatic discharge rate of I = 1 µA/cm$^2$. We observe an increase in the capacity with increased water concentration due to an increased contribution through the solution pathway. There is a nearly ~5 fold enhancement in capacity at 4000 ppm of water relative to the anhydrous case. A normalization factor for surface area of ~250 cm$^2$ is chosen such that the discharge capacity matches with that obtained from experiments.

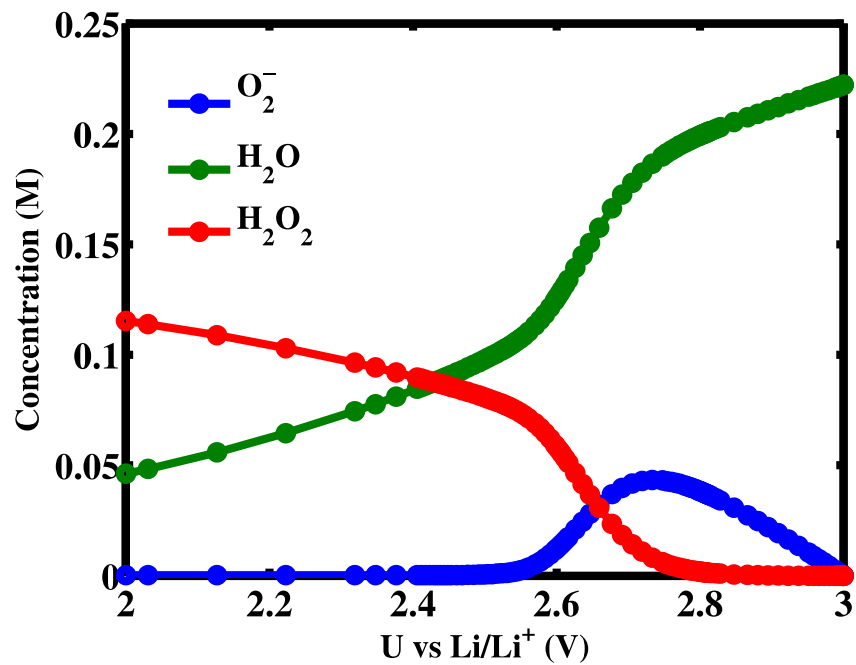

**Figure S17| Theoretically predicted concentrations in a linear sweep voltamogramm**. Concentration profiles of $O_2^-$, $H_2O$ and $H_2O_2$ as a function of potential for 4000 ppm water in the electrolyte. As observed in the galvanostatic discharge, there is an increase in the soluble $O_2^-$ concentration, which decays when the water is consumed into forming $H_2O_2$.

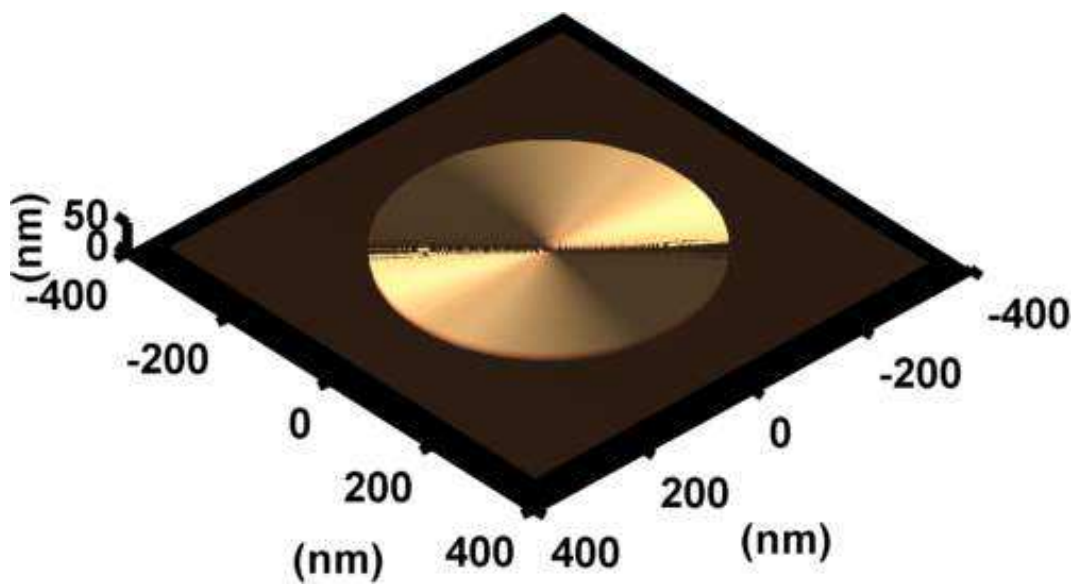

**Figure S18| Particle morphology for the high current discharge, i = 3 mA/cm$^2$** with 4000 ppm of $H_2O$. There is only a slight increase in the thickness of $Li_2O_2$, with an almost conformal coating.